\documentclass[apj]{emulateapj}
\usepackage{lscape}

\newcommand{\hii}         {\mbox{\rm \ion{H}{2}}}

\newcommand{\kms}         {km~s$^{-1}$}

\newcommand{\ha}          {\mbox{H$\alpha$}}

\def\spose#1{\hbox to 0pt{#1\hss}}
\def\lta{\mathrel{\spose{\lower 3pt\hbox{$\mathchar"218$}}
     \raise 2.0pt\hbox{$\mathchar"13C$}}}
\def\gta{\mathrel{\spose{\lower 3pt\hbox{$\mathchar"218$}}
    \raise 2.0pt\hbox{$\mathchar"13E$}}}

\shorttitle{Variable Na in a Low Extinction Type Ia SN}
\shortauthors{Simon et al.}
\slugcomment{Accepted for publication in \emph{The Astrophysical Journal}}

\begin{document}

\title{Variable Sodium Absorption in a Low-Extinction Type Ia
  Supernova\altaffilmark{1,2}}

\author{Joshua D. Simon\altaffilmark{3}, Avishay
  Gal-Yam\altaffilmark{4}, Orly Gnat\altaffilmark{5,6}, 
  Robert M. Quimby\altaffilmark{7}, 
  Mohan Ganeshalingam\altaffilmark{8},
  Jeffrey M. Silverman\altaffilmark{8}, 
  Stephane Blondin\altaffilmark{9}, Weidong Li\altaffilmark{8},
  Alexei V. Filippenko\altaffilmark{8}, J. Craig
  Wheeler\altaffilmark{10}, Robert P. Kirshner\altaffilmark{11}, 
  Ferdinando Patat\altaffilmark{9}, Peter Nugent\altaffilmark{12}, 
  Ryan J. Foley\altaffilmark{11,13}, 
  Steven S. Vogt\altaffilmark{14}, 
  R. Paul Butler\altaffilmark{15}, 
  Kathryn  M. G. Peek\altaffilmark{8},  
  Erik Rosolowsky\altaffilmark{16}, 
  Gregory J. Herczeg\altaffilmark{17}, 
  Daniel N. Sauer\altaffilmark{18}, 
  and Paolo A. Mazzali\altaffilmark{19,20,21} }

\altaffiltext{1}{Some of the data presented herein were obtained at
  the W. M. Keck Observatory, which is operated as a scientific
  partnership among the California Institute of Technology, the
  University of California, and NASA.  The Observatory was made
  possible by the generous financial support of the W. M. Keck
  Foundation.}

\altaffiltext{2}{Based in part on observations obtained with the
  Hobby-Eberly Telescope, which is a joint project of the University
  of Texas at Austin, the Pennsylvania State University, Stanford
  University, Ludwig-Maximilians-Universit{\" a}t M{\" u}nchen, and
  Georg-August-Universit{\" a}t G{\" o}ttingen.}

\altaffiltext{3}{Observatories of the Carnegie Institution of
  Washington, 813 Santa Barbara St., Pasadena, CA 91101; jsimon@ociw.edu .}

\altaffiltext{4}{Benoziyo Center for Astrophysics, Faculty of Physics,
  Weizmann Institute of Science, 76100 Rehovot, Israel;
  avishay.gal-yam@weizmann.ac.il .}

\altaffiltext{5}{Theoretical Astrophysics, California Institute of
  Technology, Mail Code 130-33, 1200 E. California Blvd., Pasadena, CA
  91125; orlyg@tapir.caltech.edu .}

\altaffiltext{6}{Chandra Fellow.}

\altaffiltext{7}{Department of Astronomy, California Institute of
  Technology, 1200 E. California Blvd., MS 105-24, Pasadena, CA 91125;
  quimby@astro.caltech.edu .}

\altaffiltext{8}{Department of Astronomy, University of California,
  Berkeley, CA 94720-3411; mganesh@astro.berkeley.edu,
  jsilverman@astro.berkeley.edu, weidong@astro.berkeley.edu
  alex@astro.berkeley.edu, kpeek@astro.berkeley.edu .}

\altaffiltext{9}{European Southern Observatory, Karl Schwarzschild Str. 2,
                 D-85748 Garching bei M{\"u}nchen, Germany;
                 sblondin@eso.org, fpatat@eso.org .}

\altaffiltext{10}{McDonald Observatory and Department of Astronomy, 
                University of Texas, Austin, TX  71782;
                wheel@astro.as.utexas.edu .}

\altaffiltext{11}{Harvard-Smithsonian Center for Astrophysics, 60
  Garden St., Cambridge, MA 02138; kirshner@cfa.harvard.edu,
  rfoley@cfa.harvard.edu .}

\altaffiltext{12}{Lawrence Berkeley National Laboratory, 1 Cyclotron
  Rd., Berkeley, CA  94720; penugent@lbl.gov .}

\altaffiltext{13}{Clay Fellow.}

\altaffiltext{14}{UCO/Lick Observatory, University of California,
  Santa Cruz, CA  95064; vogt@ucolick.org .}

\altaffiltext{15}{Department of Terrestrial Magnetism, Carnegie
  Institution of Washington, 5241 Broad Branch Rd. NW, Washington, DC
  20015; butler@dtm.ciw.edu .}

\altaffiltext{16}{Department of Physics, University of British Columbia,
                 Okanagan, British Columbia, Canada; erik.rosolowsky@ubc.ca .}

\altaffiltext{17}{Max-Planck-Institut f{\"u}r Extraterrestiche Physik,
  Postfach 1312, 85741 Garching, Germany; gregoryh@mpe.mpg.de .}

\altaffiltext{18}{Department of Astronomy, Stockholm University,
  106 91 Stockholm, Sweden; dsauer@astro.su.se .}

\altaffiltext{19}{Max-Planck-Institut f{\"u}r Astrophysik,
  Karl-Schwarzschild-Strasse 1, 85741 Garching, Germany;
  mazzali@mpa-garching.mpg.de .}

\altaffiltext{20}{Scuola Normale Superiore, Piazza dei Cavalieri 7,
  56126 Pisa, Italy .}

\altaffiltext{21}{INAF-Osservatorio Astronomico di Padova,
  Vicolo dell'Osservatorio 5, 35122 Padova, Italy .}

\begin{abstract}

Recent observations have revealed that some Type Ia supernovae exhibit
narrow, time-variable \ion{Na}{1}~D absorption features.  The origin
of the absorbing material is controversial, but it may suggest the
presence of circumstellar gas in the progenitor system prior to the
explosion, with significant implications for the nature of the
supernova progenitors.  We present the third detection of such
variable absorption, based on six epochs of high-resolution
spectroscopy of the Type Ia supernova SN~2007le from the Keck~I
telescope and the Hobby-Eberly Telescope.  The data span a time frame
of approximately three months, from 5 days before maximum light to 90
days after maximum.  We find that one component of the \ion{Na}{1}~D
absorption lines strengthened significantly with time, indicating a
total column density increase of $\sim2.5 \times 10^{12}$~cm$^{-2}$.
The data limit the typical timescale for the variability to be more
than 2 days but less than 10 days.  The changes appear to be most
prominent after maximum light rather than at earlier times when the
ultraviolet flux from the supernova peaks.  As with SN~2006X, we
detect no change in the \ion{Ca}{2}~H and K absorption lines over the
same time period, rendering line-of-sight effects improbable and
suggesting a circumstellar origin for the absorbing material.  Unlike
the previous two supernovae exhibiting variable absorption, SN~2007le
is not highly reddened ($E_{B-V} = 0.27$~mag), also pointing toward
circumstellar rather than interstellar absorption.  Photoionization
calculations show that the data are consistent with a dense
($10^{7}$~cm$^{-3}$) cloud or clouds of gas located $\sim0.1$~pc ($3
\times 10^{17}$~cm) from the explosion.  These results broadly support
the single-degenerate scenario previously proposed to explain the
variable absorption, with mass loss from a nondegenerate companion
star responsible for providing the circumstellar gas.  We also present
possible evidence for narrow H$\alpha$ emission associated with the
supernova, which will require deep imaging and spectroscopy at late
times to confirm.
\end{abstract}

\keywords{circumstellar matter --- supernovae: general --- supernovae:
  individual (SN~1999cl, SN~2006X, SN~2007le)}

\section{INTRODUCTION}
\label{intro}

The origin of Type Ia supernovae (SNe~Ia) has important implications
for understanding the physics of these dramatic explosions, the
evolution of binary stars, and the expansion of the universe.  Recent
interest in this last issue is particularly high because SNe~Ia are
the most effective distance indicators known on cosmological scales.
These objects are thus the focus of many current and future
dark-energy experiments
\citep[e.g.,][]{astier06,riess07,wv07,aldering05}.  Investigating
systematics that could affect their luminosity evolution in any way is
a prerequisite for precision cosmological studies using SNe~Ia as
calibratable standard candles.

Perhaps the most serious gap in our understanding of SNe~Ia is our
lack of knowledge about their progenitor systems.  It is widely
thought that SNe~Ia occur in close binary systems \citep{wi73}, in
which a carbon-oxygen white dwarf accretes matter from another star
until it nears the \citet{chandra} mass and its core ignites a
thermonuclear explosion that propagates outward and destroys the white
dwarf.  The identity of the companion star, however, is unknown;
possibilities range from a second white dwarf \citep[a
  double-degenerate system;][]{it84,webbink84} to a main-sequence
star, subgiant, or evolved red giant (all single-degenerate systems).
At best, these proposed progenitor systems would have an absolute
magnitude of $M_{V} \approx -3$ (for a red giant companion star near
the tip of the red giant branch), leaving them undetectable by current
telescopes for all but the closest galaxies.  Since the last known
Local Group SN~Ia occurred in 1885 \citep*{dvc85,fesen89} and there
have been no SNe~Ia observed in the Milky Way in over 400 years
\citep{rl04a,badenes06,krause08}, direct empirical constraints on the
nature of the progenitors are severely lacking.

Recent observations, however, have yielded progress on this issue
using several different approaches.  First, statistical studies of
large numbers of SN host galaxies have demonstrated a strong
relationship between Hubble type and the rate of Type Ia supernovae,
with late-type galaxies hosting as many as 20 times more SNe~Ia per
unit mass as early-type galaxies \citep[e.g.,][]{ot79, mannucci05,
  sullivan06}.  This result suggests that many SNe~Ia come from a
relatively young stellar population, although of course the SNe~Ia in
elliptical galaxies must have their origin in an older population.
These observations, combined with the SN~Ia rate as a function of
redshift, can be used to derive the delay-time distribution (the
amount of time between the formation of a stellar population and the
occurrence of SNe~Ia).  The measured delay-time distribution includes
both a prompt ($\sim10^{8}$~yr) component that is closely associated
with recent star formation and a much more extended component with a
characteristic timescale of several Gyr \citep*[][although see
  \citealt{totani08} for a conflicting view]{sb05,mannucci06}.  These
two distinct components of the SN~Ia population would seem to
naturally correspond to different mechanisms for producing Type Ia
supernovae, and perhaps different progenitor systems as well.
Supporting this idea, a number of studies have demonstrated that
SNe~Ia in early-type and late-type galaxies have different average
luminosities \citep[e.g.,][]{hamuy96,howell01,li01a,hicken09}.

Detailed studies of individual nearby SNe~Ia have also revealed
crucial clues.  \citet*{quimby07} found unusual behavior of the
velocity of the \ion{Si}{2}~$\lambda6355$ line in the otherwise normal
SN~2005hj and argued that this object might be the prototype of a
separate class of SN~Ia explosions.  \citet{rl04} obtained spectra of
stars near the center of the remnant of Tycho's supernova (SN~1572)
and claimed to identify the companion star to the SN progenitor, a
G-type subgiant, although this interpretation has proved controversial
\citep[e.g.,][]{ihara07,gh08,kerz09}.

Finally, another recent advance has been made by \citet{patat}, who
detected circumstellar material (CSM) in a SN~Ia via optical
absorption lines for the first time.  Previous searches for emission
from CSM in SNe~Ia using \ha, radio, and X-ray observations have
generally yielded only upper limits
\citep{mattila05,leonard07,panagia06,immler06,hughes07}, with the
exception of a few unusual objects whose classification as SNe~Ia has
been disputed \citep[e.g.,][but see \citealt{benetti06} and
  \citealt{trundle08} for alternative
  interpretations]{li01b,li03,hamuy03,aldering06,prieto07}.  By
obtaining high-resolution spectra of SN~2006X at multiple epochs,
\citeauthor{patat} showed that the strength of at least four distinct
components of the Na~I~D absorption lines varied with time.  The
temporal coverage of the data was rather sparse, but the variability
occurred over the time span between 2 days before maximum light and 61
days after.  Similar rapid changes in metal absorption-line profiles
with time are commonly seen in novae and provide clues to the
mass-transfer process in those systems \citep{williams08}.  Variable
Na~D absorption is also observed in the spectra of Milky Way stars
(generally on timescales of years to decades, although observations on
shorter timescales are lacking) and is usually attributed to small
interstellar clouds moving perpendicular to the line of sight
\citep[e.g.,][]{wf01}.  The lack of corresponding variability in the
\ion{Ca}{2} H \& K absorption features in SN~2006X, however, renders
this interpretation viable only with an appeal to peculiar chemical
abundances and/or an unusual geometry in the absorbing interstellar
clouds.

Accordingly, \citet{patat} concluded that the variable absorption
features originate in \emph{circumstellar} clouds in the progenitor
system that were ionized by the radiation from the supernova and
recombined over the following several weeks, with some of the clouds
then being collisionally reionized by the SN ejecta at later times.
This circumstellar gas could have originated either in the stellar
wind from the progenitor's companion star or in successive nova
eruptions.  Because \ion{Na}{1} has a much lower ionization potential
than \ion{Ca}{2}, the Na~D line profiles can change without an
accompanying effect in the Ca~H \& K lines if the ionizing radiation
has an appropriate spectrum.  The \citeauthor{patat} interpretation
has been challenged by \citet{chugai}, who showed that for typical red
giant wind densities Na~D absorption should not be detectable.
\citet{chugai} suggested instead that the absorbing material must be
located farther away from the supernova, possibly not associated with
the progenitor system at all.  Photometry and late-time low-resolution
spectroscopy of SN~2006X, which reveal at least one light echo from
the supernova, also provide support for a dusty circumstellar
environment \citep{wang08a,wang08b,crotts08}.  Taken as a whole, these
results appear to indicate a single-degenerate progenitor for
SN~2006X, perhaps with a red giant companion.

But is SN~2006X merely a unique, pathological object?  Or is it
broadly representative of a significant fraction of (or perhaps all)
SNe~Ia?  \citet{blondin08} have now identified a second supernova,
SN~1999cl, displaying variable Na~D absorption that is visible even at
low spectral resolution, but the lack of high-resolution data or
coverage of the Ca H \& K lines makes the interpretation of this
object more difficult.  The \citeauthor{blondin08} analysis
demonstrates that significant changes in the Na~D absorption profile
are relatively uncommon, with only two detections out of 31 SNe
examined.  Notably, SN~1999cl and SN~2006X are the two most heavily
reddened supernovae in the \citet{blondin08} sample, creating the
appearance of a connection between large (presumably interstellar)
reddening and variable absorption.

Only two other SNe~Ia (SN~2000cx and SN~2007af) have multi-epoch
high-resolution spectra available in the literature, and neither of
those objects exhibits any changes in the Na~D line profiles with time
\citep{patat07b,simon07}.  In this paper we present a similar data set
for the bright SN~Ia~2007le, including six high-resolution spectra
obtained between $-5$ and $+90$ days relative to maximum light.  We
use these data to test the \citet{patat} scenario, searching for
variability in the Na~D absorption features.  We describe all of our
observations, both photometric and spectroscopic, in
\S~\ref{observations}.  In \S~\ref{results}, we compare the light
curve and spectral evolution of SN~2007le with those of other SNe~Ia
and analyze the high-resolution spectra.  We discuss the implications
of the variable Na~D absorption features in \S~\ref{discussion} and
present our conclusions in \S~\ref{conclusions}.

\section{OBSERVATIONS AND DATA REDUCTION}
\label{observations}

SN~2007le was discovered by L. Monard \citep*{cbet1100} on 2007
October 13.79 (UT dates are used throughout this paper). Optical
spectra obtained two nights later showed that the object was a Type Ia
supernova at least one week before maximum light, and featuring
high-velocity ejecta expanding at $\sim$16,000~\kms\ \citep{cbet1101}.
The host galaxy of the supernova is NGC~7721, an Sc galaxy with a
heliocentric recession velocity of 2015~\kms\ \citep{koribalski04}.

\subsection{High-Resolution Spectroscopy}

Our high-resolution observing campaign for SN~2007le began on 2007
October 20 with the High Resolution Echelle Spectrometer
\citep[HIRES;][]{vogt94} on the Keck~I 10~m telescope.  Over the
following three months, we obtained a total of five HIRES spectra of
the supernova.  The seeing was often poor ($\gta1\arcsec$) during the
observations, and the overall conditions varied significantly over the
course of the many observing runs.  The data were obtained with a
range of different spectrograph setups, including spectra with both
the blue and red cross-dispersers, and achieved signal-to-noise ratios
(S/N) ranging between 11 and 113 per pixel.  Most of the spectra used
a $7.0\arcsec \times 0.86\arcsec$ slit, yielding a spectral resolution
of $R \approx 52,000$, but one spectrum was obtained with a wider
($7.0\arcsec \times 1.15\arcsec$) slit and $R \approx 42,000$.  All of
the HIRES spectra cover the Ca~H \& K, Na~D, and \ha\ lines.  We
reduced the Keck data using the IDL data reduction package for HIRES
(version 2.0) developed by J.~X.~Prochaska and collaborators
(Bernstein et al., in prep.).\footnote{Documentation and code for this
  reduction package are available at
  http://www.ucolick.org/$\sim$xavier/HIRedux/index.html.}

We also obtained one spectrum of SN~2007le with the High-Resolution
Spectrograph \citep[HRS;][]{hrs} on the Hobby-Eberly Telescope (HET)
on 2007 November 4.  The spectrograph was in its $R = 60,000$ mode,
with a 2\arcsec-diameter fiber and the 316 line/mm grating centered at
6948~\AA, providing nearly complete wavelength coverage from 5100 to
8850~\AA.  We obtained 3 spectra totaling 3000~s of exposure time and
reached a combined S/N of 16 per pixel.  The HRS data were reduced in
IRAF\footnote{IRAF is distributed by the National Optical Astronomy
  Observatories, which are operated by the Association of Universities
  for Research in Astronomy, Inc., under cooperative agreement with
  the National Science Foundation.} with the {\sc echelle} package
using standard procedures.  A comprehensive summary of all the
high-resolution data is presented in Table~\ref{obstable}.

\begin{deluxetable*}{lllccccc}
\tablewidth{0pt}
\tabletypesize{\scriptsize}
\tablecolumns{8}
\tablecaption{High-Resolution Spectroscopy Observing Log}
\tablehead{
\colhead{Telescope} & \colhead{Instrument} & 
\colhead{UT Date} & \colhead{Supernova Epoch\tablenotemark{a}} &
\colhead{Exposure Time} & 
\colhead{S/N\tablenotemark{b}} & 
\colhead{Spectral Resolution\tablenotemark{c}} & 
\colhead{Wavelength Range} \\
\colhead{} & \colhead{} & 
\colhead{} & 
\colhead{} & \colhead{[s]} & 
\colhead{} & 
\colhead{($\lambda/\Delta\lambda$)} & \colhead{[\AA]} } 
\startdata 
Keck I  & HIRESb & 2007 October 20.36 & day $-5\phn$ & 7200 & 113 & 52,000 & 3840--6733 \\
Keck I  & HIRESr & 2007 October 25.34 & day $0\phn\phn\phd$ & 900  & 71  & 41,700 & 3460--7989 \\
HET     & HRS    & 2007 November 4.15 & day $+10$ & 3000 & 16  & 60,000 & 5100--8850 \\
Keck I  & HIRESb & 2007 November 6.39 & day $+12$ & 4500 & 76  & 47,600 & 3845--6649 \\
Keck I  & HIRESr & 2008 January 17.23 & day $+84$ & 3300 & 46  & 54,000 & 3870--8365 \\
Keck I  & HIRESr & 2008 January 23.21 & day $+90$ & 1200 & 11  & 54,000 & 3900--8360 \\
\enddata
\tablenotetext{a}{Relative to maximum light.}
\tablenotetext{b}{Signal-to-noise ratio measurements are per pixel at
  the wavelength of the redshifted Na~D lines (5934.5--5936.5~\AA).}
\tablenotetext{c}{The spectral resolution was determined by measuring
  the FWHM of emission lines in the comparison-lamp spectra and telluric
  absorption lines in the SN and telluric standard star spectra.}
\label{obstable}
\end{deluxetable*}

\subsection{Imaging and Low-Resolution Spectroscopy}

SN~2007le was the target of extensive photometric follow-up
observations with the 0.76~m Katzman Automatic Imaging Telescope
\citep[KAIT;][]{li00,filippenko01}, continuing for approximately 3
months until the supernova went into conjunction with the Sun.  We
obtained post-explosion images in 2008 August after the SN had faded
in order to subtract the host-galaxy light.  We used the {\sc daophot}
package \citep{stetson87} in IRAF to perform point-spread function
photometry of SN~2007le relative to various field stars in the KAIT
images, which were calibrated on 5 photometric nights with KAIT and
the Nickel 1~m telescope at Lick Observatory.

We also obtained low-resolution spectra of SN~2007le with the Low
Resolution Imaging Spectrometer \citep[LRIS;][]{oke95} on the Keck~I
telescope on 2007 October 15, October 16, November 11, November 12,
and December 12, and with the Kast spectrograph \citep{miller93} on
the 3~m Shane telescope at Lick Observatory on 2007 November 2,
November 18, and December 1.  These data were reduced in IRAF and IDL
following normal procedures \citep[for details,
  see][]{foley03,matheson00,horne86}.
%

\section{DATA ANALYSIS AND RESULTS}
\label{results}

\subsection{Light Curves and Low-Resolution Spectra}

\begin{figure}[t!]
\epsscale{1.24}
\plotone{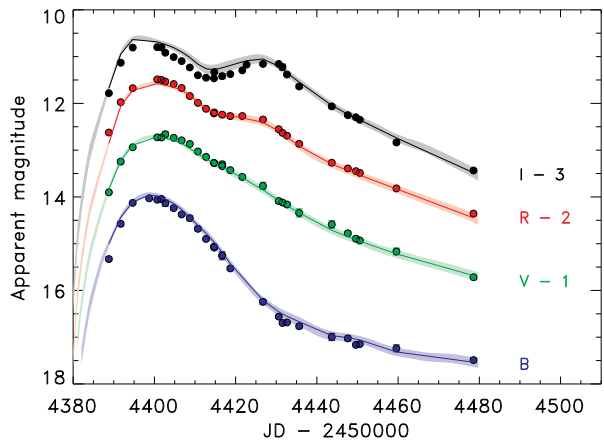}
\caption{$BVRI$ light curves (filled symbols) of SN~2007le, from KAIT
  and Lick Nickel 1~m data.  Photometric uncertainties are indicated
  by the plotted error bars, which in most cases are smaller than the
  displayed data points.  The solid lines represent the MLCS2k2 fits
  to the data and the shaded bands represent the uncertainties on the
  fits.}
\label{lightcurves}
\end{figure}

We display $BVRI$ light curves of SN 2007le in Figure
\ref{lightcurves}.  We fitted the photometric data with the latest
version of the multicolor light-curve shape method
\citep*[MLCS2k2;][]{jrk07} to determine the parameters of the
supernova.  We find that the time of $B$-band maximum was 2007 October
25.65 (JD = 2,454,399.15), with an uncertainty of 0.09~day.  The
derived line-of-sight extinction to the SN is $A_{V} = 0.71 \pm
0.06$~mag, with an extinction law of $R_{V} = 2.56 \pm 0.22$ (the
Milky Way foreground reddening is 0.033~mag, corresponding to $A_{V} =
0.11$~mag; \citealt*{sfd98}).  The distance modulus to SN~2007le is $m
- M = (32.35 - 5 \log[{\rm H}_{0}/(70~{\rm km}~{\rm s}^{-1}~{\rm
    Mpc}^{-1})]) \pm 0.06$~mag, giving the SN an absolute magnitude of
$M_{V} = (-19.34 + 5\log{{\rm H}_{0}/70}) \pm 0.09$.  The
luminosity/light-curve-shape parameter is $\Delta = -0.14 \pm 0.02$,
and the MLCS2k2 $\chi^{2}$ value of 163.0 for 106 degrees of freedom
indicates an acceptable fit (much of the $\chi^{2}$ comes from the
$I$-band data, which significantly differ from the fit).

\begin{figure}[t!]
\epsscale{1.24}
\plotone{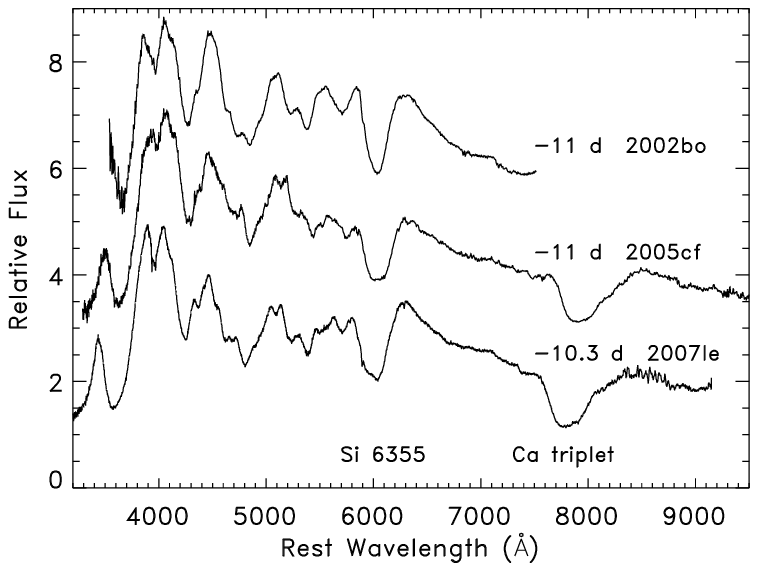}
\caption{Low-resolution spectrum of SN 2007le 10.3 days before maximum
  light (lower curve), compared to scaled spectra of SN~2005cf
  (middle, data from \citealt{wang09}) and SN~2002bo (top, data from
  \citealt{benetti04}) at similar epochs.  The spectral features
  match closely between the three supernovae.  The SN~2005cf and
  SN~2002bo data have been dereddened by $A_{V} = 0.58$~mag
  \citep{wang09} and $A_{V} = 1.0$~mag \citep{eliasrosa08},
  respectively, and then re-reddened by $A_{V} = 0.71$~mag to match
  the extinction of SN~2007le.}
\label{lowresspectra}
\end{figure}

In Figure \ref{lowresspectra} we show the spectrum of SN~2007le at an
epoch of 10.3 days before maximum light.  Analysis with the Superfit
spectral fitting code of \citet{howell05} indicates that SN~2007le is
most similar to spectra of the highly reddened type Ia SN~2002bo at 11
days before maximum from \citet{benetti04}, while the SNID package
\citep{bt07} finds a best fit to early spectra of the ``golden
standard'' type Ia SN~2005cf \citep{wang09}.  Spectra of SN~2002bo and
SN~2005cf were included in the databases used by both fitting
packages; the differences between the two lie in their fitting methods
and treatment of color information (see \citealt{bt07}), but since
SN~2002bo and SN~2005cf were similar events in many respects (e.g.,
line velocities) we do not regard these fit results as a significant
disagreement.  Like SN~2002bo (and SN~2006X; \citealt{wang08a}),
SN~2007le exhibits a high-velocity component to the \ion{Ca}{2}
near-infrared triplet absorption lines, although such features are
very common in SNe~Ia that are observed at sufficiently early times
\citep{mazzali05b}.  Both SN~2006X and SN~2002bo were highly polarized
\citep{wang06,wang07,ww08}, suggesting that polarization measurements
of SN~2007le might be very interesting as well.

In order to classify SN~2007le among the various known subgroups of
Type Ia supernovae, we analyzed the spectra according to the
prescriptions of \citet{benetti05} and \citet{branch06,branch09}.
The widths of the absorption features near 5750~\AA\ and
6100~\AA\ correspond to the broad line (BL) group of
\citeauthor{branch09}, but the equivalent widths are not too far from
those of the core normal group.  We measured a velocity gradient in
the \ion{Si}{2}~$\lambda$6355 line of $83\pm3$~\kms~day$^{-1}$,
placing SN~2007le marginally within the high velocity gradient (HVG)
class of \citet{benetti05}, although this gradient is not very much
larger than is seen in some low velocity gradient SNe.  SN~2006X and
SN~2002bo both exhibited significantly larger gradients.

\subsection{Removal of Telluric Absorption Features}
\label{telluric_lines}

The host-galaxy Na~D absorption lines of SN~2007le are located at
observed wavelengths between 5930 and 5940~\AA.  This region of the
spectrum unfortunately contains a number of telluric absorption lines
from water molecules \citep*[e.g.,][]{moore66}.  Before searching for
variations in the intrinsic supernova line profiles, we therefore must
remove the telluric features, which will change with atmospheric
conditions \citep[e.g.,][]{wh88,matheson01}.

On every night except that of the second-epoch observations (day~0,
2007 October 25) we obtained at least one spectrum of a hot, rapidly
rotating star to serve as a telluric standard.  These spectra were
reduced and normalized to a flat continuum level in the same manner as
the supernova data.  Because the telluric standards do not contain any
atmospheric features in the region of interest, any absorption seen
must necessarily be of telluric origin.  From the telluric standard
observations we constructed telluric absorption line model spectra
using a list of known telluric lines.\footnote{We use a model rather
  than simply the observed standard star spectrum in order to remove
  any effects from imperfectly fitting the continuum of the standard
  star and because the S/N of the standard spectra is in some cases
  not too much higher than the S/N of the supernova spectrum.}  Pixels
in the model spectra that were at the wavelength of telluric lines
were set equal to the value of those pixels in the telluric standard
star spectra, and all other pixels were set to unity.  We then
selected the telluric model obtained at the most similar airmass and
time for each supernova spectrum.\footnote{For the second-epoch
  observations where we lack a telluric standard from the same night,
  we use a telluric standard from the night of 2007 October 20, which
  had the most similar atmospheric water vapor content to 2007 October
  25 according to the Mauna Kea Weather Center forecasts (available on
  the internet) for the nights on which we observed.}  The telluric
models were further adjusted to match the airmass of the supernova
observations with the following scaling: model$_{adjusted} = $
(model$_{original}$)$^{b}$, where $b = $
airmass$_{SN}$/airmass$_{standard}$.  Finally, we divided the reduced
supernova spectra by their respective telluric absorption models to
produce clean SN spectra.  The strength of the absorption lines that
overlap with the wavelength of the host-galaxy Na~D absorption in the
telluric models ranges from a few percent to 10\%, not nearly large
enough to account for the absorption profile changes described in \S
\ref{na}.

\subsection{Sodium D Absorption Lines at High Resolution}
\label{na}

\begin{figure*}[t!]
\epsscale{1.24}
\plotone{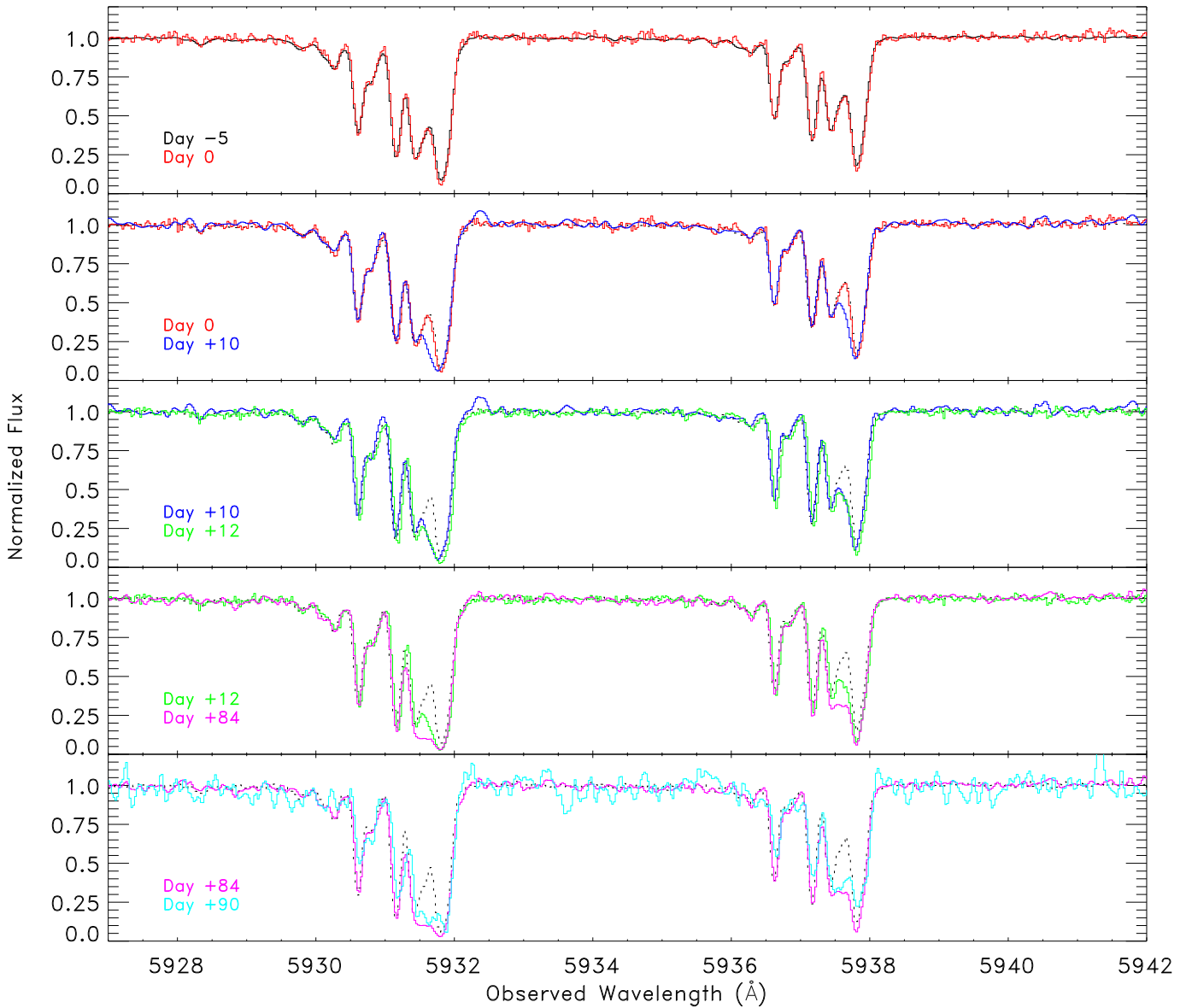}
\caption{High-resolution observations of the Na~D absorption lines in
  the spectrum of SN~2007le near the host-galaxy velocity.  The top
  panel shows the initial HIRES spectrum (day~$-5$) in black and the
  second-epoch spectrum from day~0 later in red.  In each subsequent
  panel the initial spectrum is plotted as a dotted black curve.  The
  second panel compares the day~0 (red) and day~$+10$ (blue) spectra
  to each other, and then each following panel does the same for the
  next neighboring pair of observations.  All of the spectra shown in
  each panel are smoothed to the lowest resolution spectrum displayed
  in that panel (which means that the effective resolution changes
  somewhat from panel to panel).  At least one component of the Na
  absorption near 5931.5~\AA\ in the D$_{2}$ line clearly strengthens
  with time. }
\label{highresspectra}
\end{figure*}

We display the high-resolution spectra of SN~2007le around the
host-galaxy Na~D lines in Figure~\ref{highresspectra}.  At least one
component of the absorption (near 5931.5~\AA\ and 5937.5~\AA\ in the
D$_{2}$ and D$_{1}$ lines, respectively) strengthens over the course
of the observations, increasing from a total depth of $\sim0.6$ in the
first spectrum to nearly saturated three months later.\footnote{In
  Figure~\ref{highresspectra}, it appears that the final (day~+90)
  spectrum shows additional changes, with \emph{all} of the Na
  absorption features weakening compared to the previous epochs.
  Given the complete lack of variations in all but one of these
  components through the day~+84 spectrum obtained just six days
  earlier, the very poor observing conditions under which the day~+90
  spectrum was acquired (bad seeing, high humidity, and bright sky
  because the supernova set before the end of twilight), and the low
  S/N of the spectrum, we suspect that these changes indicate a
  systematic problem with the data rather than true physical changes.
  We therefore exclude the day~+90 spectrum from the remainder of our
  analysis.}  The full absorption profile is complex, with $\sim7$
distinct velocity components visible.  Fitting the absorption with a
set of Gaussians reveals that at least 2 additional components blended
with the strongest absorption features are needed in order to remove
significant residuals, and even then the fits to the high-S/N spectra
are not statistically satisfactory.  Fits of multiple blended
Gaussians do not produce unique results, however, so we cannot study
the physical conditions in the variable absorption component(s)
accurately with this technique.

Instead, we first integrate directly over the entire absorption
profile to determine the equivalent width (EW) of the D$_{1}$ and
D$_{2}$ lines in each spectrum.  The measured equivalent widths are
listed in columns (2) and (3) of Table \ref{na_ew}.  Over the course
of the observations, both lines show an increase in the EW of slightly
more than 100 m\AA.  The total EWs in the day~$-5$ spectrum are $894
\pm 2$~m\AA\ and $649 \pm 2$~m\AA\ for the D$_{2}$ and D$_{1}$ lines,
respectively.  By the final spectrum at day~$+84$, the EWs are $1006
\pm 5$~m\AA\ and $766 \pm 5$~m\AA.  For the four blue-most components
of the absorption, we rule out any variation at high significance; the
EWs of those components agree in all six epochs to a level of a few
m\AA.  To reduce the uncertainty in the estimate of the change in EW,
we can therefore leave these components out of the integration and sum
the absorption starting from the peak at 5931~\AA\ (5937~\AA\ for the
D$_{1}$ line) instead of from the blue edge of the absorption at
5929.5~\AA\ (5935.5~\AA).  With this smaller wavelength range we find
a total increase in EW between day~$-5$ and day~$+84$ of $106 \pm
5$~m\AA\ and $105 \pm 5$~m\AA\ for the D$_{2}$ and D$_{1}$ lines (last
two columns of Table \ref{na_ew}).

\begin{deluxetable}{lllll}
\tablewidth{0pt}
\tabletypesize{\scriptsize}
\tablecolumns{5}
\tablecaption{Na D Equivalent Widths}
\tablehead{
  \multicolumn{1}{c}{} & \multicolumn{2}{c}{Full Profile}  & 
  \multicolumn{2}{c}{Red Side} \\
  \multicolumn{1}{c}{} & 
  \multicolumn{2}{c}{$\overline{\phm{SpanningSpannin}}$} &
  \multicolumn{2}{c}{$\overline{\phm{SpanningSpannin}}$} \\
\colhead{Epoch} & \colhead{D$_{2}$ EW} & 
\colhead{D$_{1}$ EW} & \colhead{D$_{2}$ EW} & 
\colhead{D$_{1}$ EW} \\
\colhead{} & \colhead{[m\AA]} & 
\colhead{[m\AA]} & \colhead{[m\AA]} & 
\colhead{[m\AA]}}
\startdata 
day $-5$ & $\phn894 \pm 3$  & $649 \pm 3$  & $653 \pm 2$  & $498 \pm 2$  \\
day $0$ & $\phn883 \pm 5$  & $661 \pm 5$  & $655 \pm 3$  & $512 \pm 3$  \\
day $+10$ & $\phn906 \pm 20$ & $695 \pm 20$ & $670 \pm 12$ & $537 \pm 12$ \\
day $+12$ & $\phn949 \pm 4$  & $702 \pm 5$  & $711 \pm 3$  & $549 \pm 3$  \\
day $+84$ & $1006 \pm 7$ & $766 \pm 7$  & $759 \pm 4$  & $603 \pm 4$  \\
\enddata
\label{na_ew}
\end{deluxetable}

It is worth noting that all of the Na~D absorption features, including
the variable component, are blueshifted relative to the strongest
absorption component at 5931.8~\AA\ (2130~\kms), which we presume
corresponds to the local interstellar medium (ISM) velocity (see \S
\ref{molecules} and \ref{ha}).  In SN~2006X and SN~1999cl, the
variable absorption was also blueshifted compared to the ISM
absorption \citep{patat,blondin08}, perhaps providing a clue as to the
origin of the material responsible for the varying absorption.

\begin{figure}[t!]
\epsscale{1.24}
\plotone{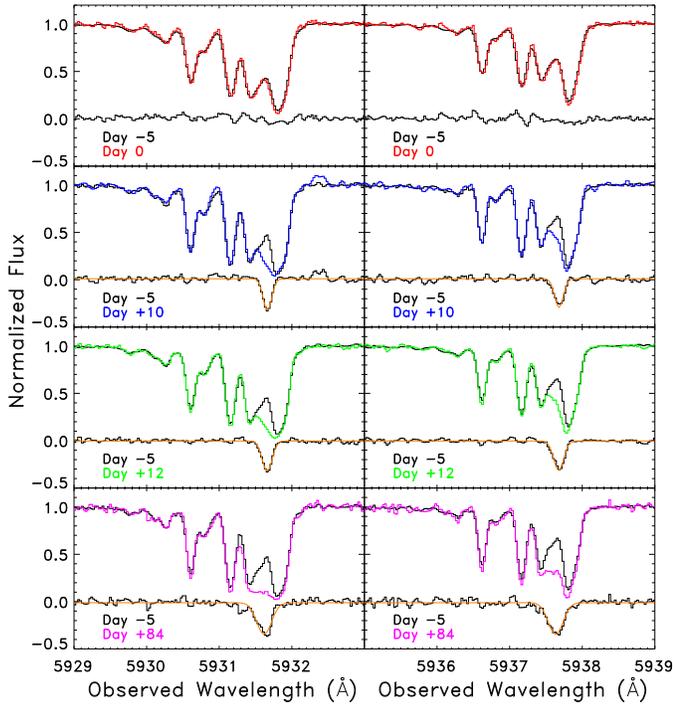}
\caption{Differences in the Na~D lines between later epochs and the
  first spectrum.  The left panels show the D$_{2}$ lines and the
  right panels the D$_{1}$ lines.  The top curves in each panel are
  the observed spectra, and the black line at the bottom of each panel
  is the difference between the two spectra plotted in that panel.
  The orange curves are Gaussian fits to the difference spectra.
  There are no significant differences between the first two spectra
  (at and before maximum light), but the later spectra all show
  additional absorption at 5931.7 and 5937.7~\AA.}
\label{diffspec}
\end{figure}

Having determined the EW of the variable absorption, we would now like
to constrain the column density and line width of the absorbing gas.
To isolate the changing component of the absorption profile, we
examine the difference between each spectrum and the first one.  These
difference spectra are displayed in Figure \ref{diffspec}.  At day~0
(maximum light), we do not detect any obvious difference signal from 5
days earlier, despite the possible small increase in EW reported for
the D$_{1}$ line in Table \ref{na_ew}.  Between day~$-5$ and
day~$+10$, however, a clear Gaussian line emerges in the difference
spectrum (Fig. \ref{diffspec}, second row from the top).  As expected
from the EW measurements above, the strength of this line increases
through day~$+84$, and we also find that the Doppler parameter of the
line grows from $4.2\pm0.7$~\kms\ in the (day~$+10$ minus day~$-5$)
spectrum to $10.7\pm0.2$~\kms\ in the (day~$+84$ minus day~$-5$)
spectrum (see Table \ref{diff_ew}).  The line in the (day~$+10$ minus
day~$-5$) difference spectrum appears close to symmetric, but stronger
hints of asymmetry in the later difference spectra suggest that some
of this increase in line width is resulting from the addition of a
second blended absorption component.

\begin{deluxetable*}{lcclc}
\tablewidth{0pt}
\tabletypesize{\scriptsize}
\tablecolumns{5}
\tablecaption{Na D Difference Spectra Fits}
\tablehead{
\colhead{Epochs} & \colhead{D$_{2}$ EW change} & 
\colhead{D$_{1}$ EW change} & \colhead{D$_{2}$ velocity} & 
\colhead{D$_{2}$ Doppler parameter} \\
\colhead{} & \colhead{[m\AA]} & 
\colhead{[m\AA]} & \colhead{[\kms]} & 
\colhead{[\kms]}}
\startdata 
day $+10$ $-$ day $-5$ & $50 \pm 4$  & $45 \pm 5$  & $2123.1 \pm 0.2$  & $\phn4.2 \pm 0.7$ \\
day $+12$ $-$ day $-5$ & $55 \pm 1$  & $54 \pm 2$  & $2122.9 \pm 0.1$  & $\phn5.3 \pm 0.2$ \\
day $+84$ $-$ day $-5$ & $85 \pm 2$  & $85 \pm 2$  & $2120.8 \pm 0.1$  & $10.7 \pm 0.2$ \\

\enddata
\label{diff_ew}
\end{deluxetable*}

\subsection{Column Density of the Variable Component}
\label{conditions}

\begin{figure}[t!]
\epsscale{1.24}
\plotone{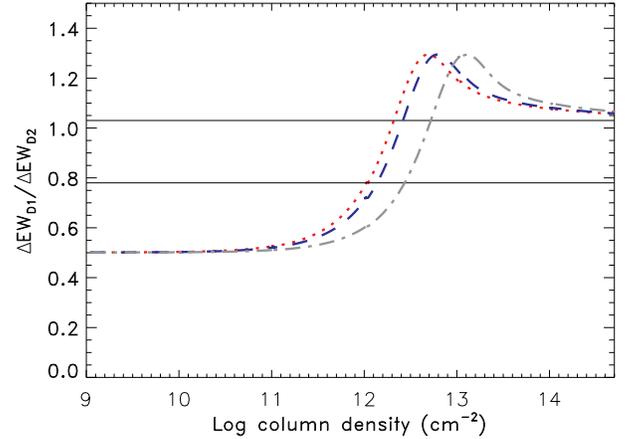}
\caption{Ratio of the change in the EW of the \ion{Na}{1} D$_{1}$
line to that of the D$_{2}$ line for a small increase in the absorbing
column as a function of column density.  The dotted red curve
represents the EW ratio for a Doppler parameter of $b = 4.2$~\kms,
appropriate for the (day~$+10$ minus day~$-5$) spectrum, the dashed
blue curve for a Doppler parameter of $b = 5.3$~\kms, appropriate for
the (day~$+12$ minus day~$-5$) spectrum, and the dot-dashed gray curve
for a Doppler parameter of $b = 10.7$~\kms, appropriate for the
(day~$+84$ minus day~$-5$) spectrum.  The horizontal solid lines
delineate the 1~$\sigma$ constraints placed by the observed EW ratios,
$0.78 \le \Delta\mbox{EW}_{\rm{D}_{1}}/\Delta\mbox{EW}_{\rm{D}_{2}}
\le 1.03$.  The observed ratio of the EW changes can only be obtained
over a narrow range of column densities from $10^{12}$~cm$^{-2}$ to
$10^{12.7}$~cm$^{-2}$. }
\label{ew_ratio}
\end{figure}

We see in Table \ref{diff_ew} that the change in the EW of the D$_{2}$
line is very similar to the change in the EW of the D$_{1}$ line at
all epochs.  At 1~$\sigma$ significance, $0.78 \le
\Delta\mbox{EW}_{\rm{D}_{1}}/\Delta\mbox{EW}_{\rm{D}_{2}} \le 1.03$
over the full set of difference spectra.  Following \citet[][Equations
  3-49 to 3-51]{spitzer78}, we can use a curve-of-growth analysis to
determine the observed EW for any combination of column density and
line width.  Using the derived Doppler parameters for the difference
spectra given in \S~\ref{na}, we calculate the change in EW that would
be observed for a small increase in column density for all column
densities between $10^{9}$~cm$^{-2}$ and $10^{14.5}$~cm$^{-2}$.  For a
given line width, these calculations demonstrate that increasing the
equivalent width of the D$_{2}$ and D$_{1}$ lines by the same amount
is only possible for a narrow range of column densities, as
illustrated in Figure \ref{ew_ratio}.  For $b = 4.2$~\kms, we find
that $12.0 \le \log{N_{\rm Na~I}} \le 12.3$, for $b = 5.3$~\kms, we
find that $12.1 \le \log{N_{\rm Na~I}} \le 12.4$, and for $b =
10.7$~\kms, we find that $12.4 \le \log{N_{\rm Na~I}} \le 12.7$.  We
therefore conclude that the column density of the varying component of
the absorption is $N_{\rm Na~I} \approx 2.5 \times 10^{12}$~cm$^{-2}$.
Depending on the Doppler parameter, such a column density corresponds
to a total Na D$_{2}$ EW of 200--300~m\AA\ for this component,
consistent with the results shown in Table \ref{na_ew}.

It is evident from Figure~\ref{ew_ratio} that the column densities
indicated by the first two difference spectra (blue dashed and red
dotted curves) are not consistent with the column density preferred by
the third difference spectrum (gray dot-dashed curve).  We suspect
that the reason for this inconsistency is that the third difference
spectrum contains multiple blended components, making the measured
line width larger than the true line width as noted in \S~\ref{na}.
However, without a higher-resolution spectrum that separates the two
components, we do not have conclusive evidence in favor of this
interpretation.

\subsection{Calcium H\&K Absorption Lines at High Resolution}
\label{ca}

\begin{figure}[t!]
\epsscale{1.24}
\plotone{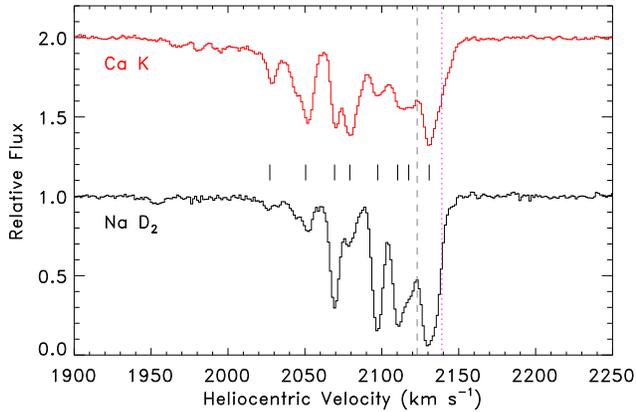}
\caption{Comparison between the host-galaxy absorption profile in the
  \ion{Ca}{2} K (upper spectrum, in red) and \ion{Na}{1} D$_{2}$
  (lower spectrum, in black) lines, both from five days before maximum
  light.  The vertical dashed gray line is at the central velocity of
  the variable Na absorption component.  Each Na absorption component,
  including the variable one, has a visible Ca counterpart (and vice
  versa), with the possible exception of some very weak blueshifted
  features between 1950 and 2000~\kms.  The relative strengths of the
  Na and Ca lines, however, vary significantly from one component to
  the next.  The Ca lines also appear to have much more prominent
  wings than the Na absorption.  The vertical ticks mark the
  velocities of individual absorption components inferred from
  Gaussian fits to the Na lines, and the vertical dotted magenta line
  indicates the velocity of the \ha\ emission line discussed in
  \S~\ref{ha}. }
\label{na_vs_ca}
\end{figure}

The Keck/HIRES spectra extend far enough to the blue that we also
detect \ion{Ca}{2} H \& K absorption lines from both the Milky Way and
the host galaxy, although the S/N of the day~$+90$ spectrum in the
blue is too low to be useful.  We compare the host-galaxy absorption
profile in Ca K and Na D$_{2}$ in Figure \ref{na_vs_ca}, where a close
correspondence between the absorption components in each species is
visible.

\begin{figure*}[t!]
\epsscale{1.24}
\plotone{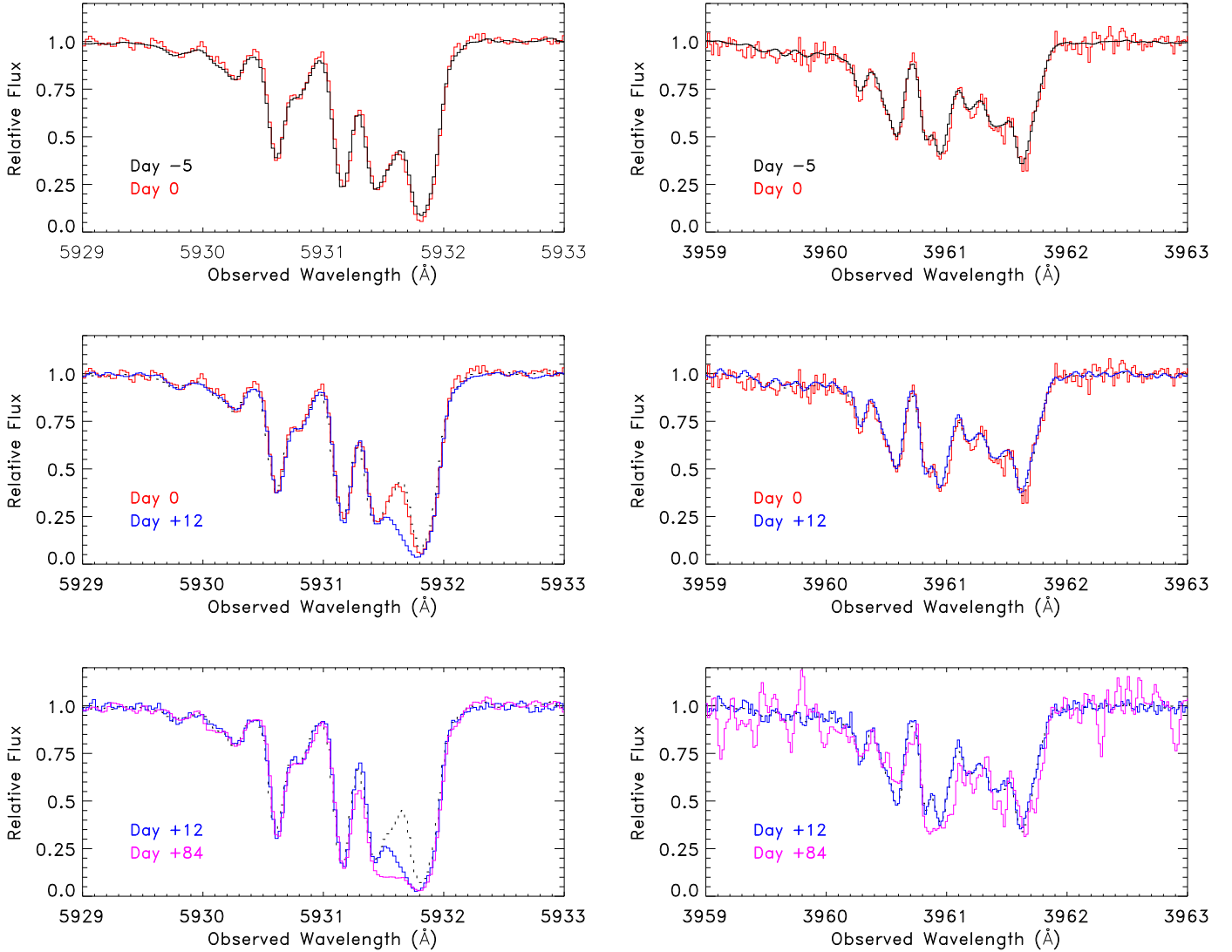}
\caption{Comparison of the Na~D$_{2}$ (left panels) and Ca~K (right
  panels) line profiles as a function of time.  The top row includes
  spectra from day~$-5$ and day~0, the middle row includes spectra
  from day~0 and day~$+12$, and the bottom row includes spectra from
  day~$+12$ and day~$+84$.  The dotted black curve in the middle and
  lower panels represents the day~$-5$ spectrum.  Over the time period
  when significant changes are visible in the Na lines there are no
  detectable variations in the Ca lines.}
\label{highresspectra_ca}
\end{figure*}

Unlike the Na lines, though, we detect no statistically significant
changes in the Ca absorption profile with time.  Even nearly three
months after maximum light, the profile shape (Figure
\ref{highresspectra_ca}) and total EW (Table \ref{ca_ew}) match those
from before maximum light within the uncertainties.  However, while
changes in the Ca EW as large as those seen in the Na lines can be
ruled out, the low S/N of the late-time spectra prevents us from being
able to place strong constraints on smaller variations.

\begin{deluxetable}{lllll}
\tablewidth{0pt}
\tabletypesize{\scriptsize}
\tablecolumns{5}
\tablecaption{Ca H \& K Equivalent Widths}
\tablehead{
  \multicolumn{1}{c}{} & \multicolumn{2}{c}{Full Profile}  & 
  \multicolumn{2}{c}{Red Side} \\
  \multicolumn{1}{c}{} & 
  \multicolumn{2}{c}{$\overline{\phm{SpanningSpannin}}$} &
  \multicolumn{2}{c}{$\overline{\phm{SpanningSpannin}}$} \\
\colhead{Epoch} & \colhead{K EW} & 
\colhead{H EW} & \colhead{K EW} & 
\colhead{H EW} \\
\colhead{} & \colhead{[m\AA]} & 
\colhead{[m\AA]} & \colhead{[m\AA]} & 
\colhead{[m\AA]}}
\startdata 
day $-5$ & $653 \pm 2$  & $382 \pm 2$  & $301 \pm 1$  & $174 \pm 1$  \\
day 0 & $677 \pm 9$  & $397 \pm 9$  & $308 \pm 4$  & $178 \pm 5$  \\
day $+12$ & $629 \pm 6$  & $388 \pm 6$  & $294 \pm 3$  & $186 \pm 3$  \\
day $+84$ & $749 \pm 32$ & $413 \pm 28$ & $343 \pm 16$  & $153 \pm 15$  \\

\enddata
\label{ca_ew}
\end{deluxetable}

\subsection{Other Absorption Features}
\label{molecules}

In addition to the Ca~H \& K and Na~D lines, we also searched the
spectra for the additional absorption features identified by
\citet{patat} and \citet{cp08} in SN~2006X.  In the highest S/N
(day~$-5$) spectrum, we detect the CH$^{+}$ 3957.70~\AA\ and
4232.55~\AA\ lines at a velocity of 2130~\kms, exactly matching the
velocity of the strongest Na~D absorption.  Diffuse interstellar bands
are also visible at 5780~\AA\ and 6283~\AA\ (rest wavelengths).  We do
not detect \ion{Ca}{1} at 4226.73~\AA, CH at 4300.30~\AA, the CN
vibrational band, or \ion{K}{1} at 7699~\AA.

\subsection{\ha\ Emission}
\label{ha}

All six of our high-resolution spectra cover the expected wavelength
of the redshifted \ha\ line.  However, the fiber spectrum from the HET
is not suitable for investigating possible \ha\ emission from the SN
because of the lack of local sky subtraction, so we only consider the
five high-resolution Keck spectra in this section.

\begin{figure*}[t!]
\epsscale{1.24}
\plotone{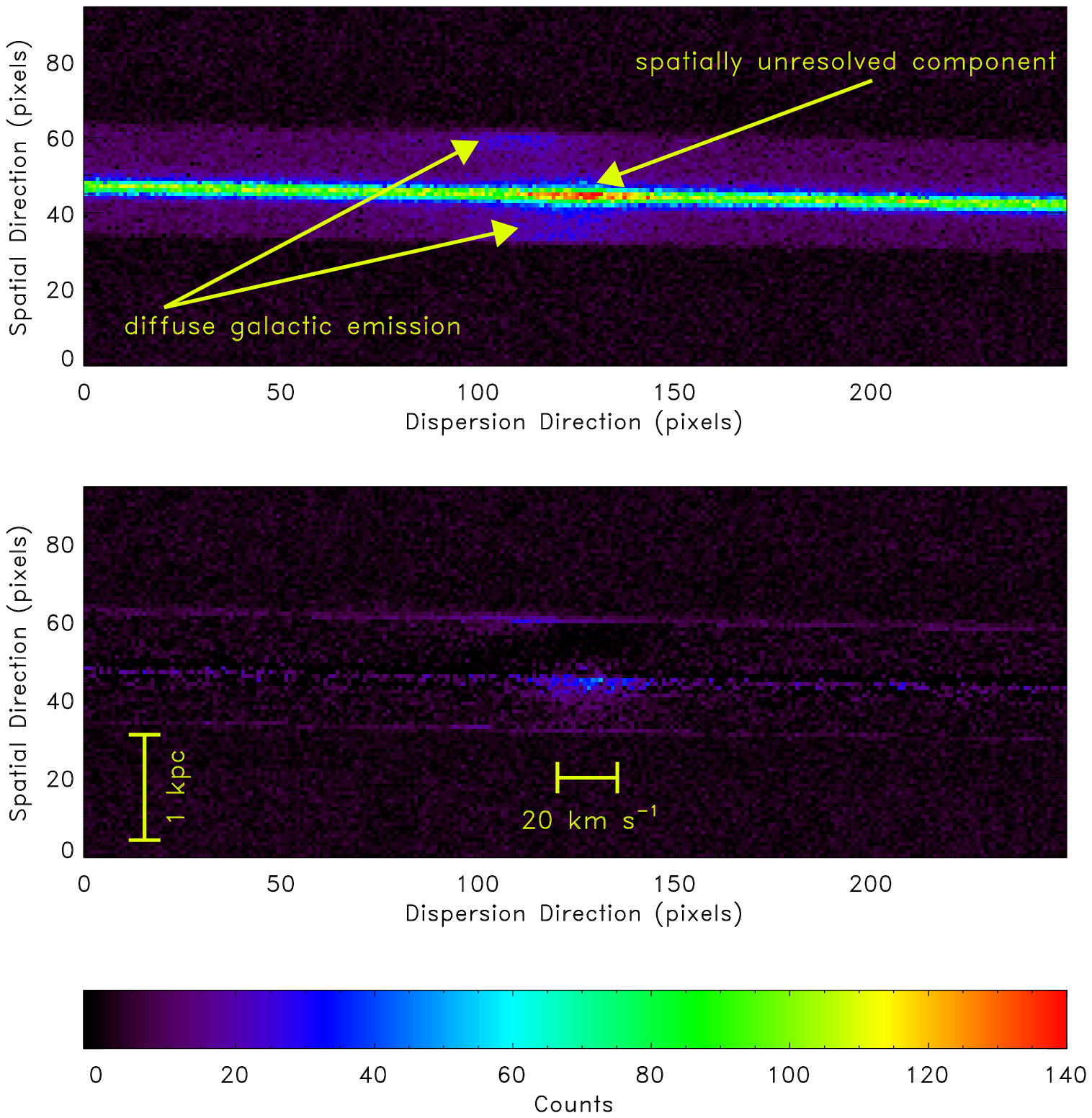}
\caption{Two-dimensional spectra of SN~2007le around the redshifted
  \ha\ line.  These data are from day~$+84$, where the S/N of the
  \ha\ detection is highest.  The top panel shows a portion of the
  echelle order containing \ha\ in the coadded raw frames (the only
  processing was to remove cosmic rays).  Diffuse \ha\ emission is
  visible across the slit, with a velocity that changes from one side
  to the other, but an additional distinct component is present right
  on top of the SN continuum.  In the lower panel, the continuum
  emission and the average sky light across the slit have been
  subtracted.  The sky fitting is not perfect, since the diffuse
  \ha\ was not symmetrically distributed around the supernova, but it
  is clear that there is excess \ha\ light at the position of the SN
  compared with the surrounding areas of the host galaxy.
  Fully reduced and extracted one-dimensional spectra of this emission
  are displayed in Figure \ref{ha_1dspec}.}
\label{ha_2dspec}
\end{figure*}

\begin{figure}[th!]
\epsscale{1.24}
\plotone{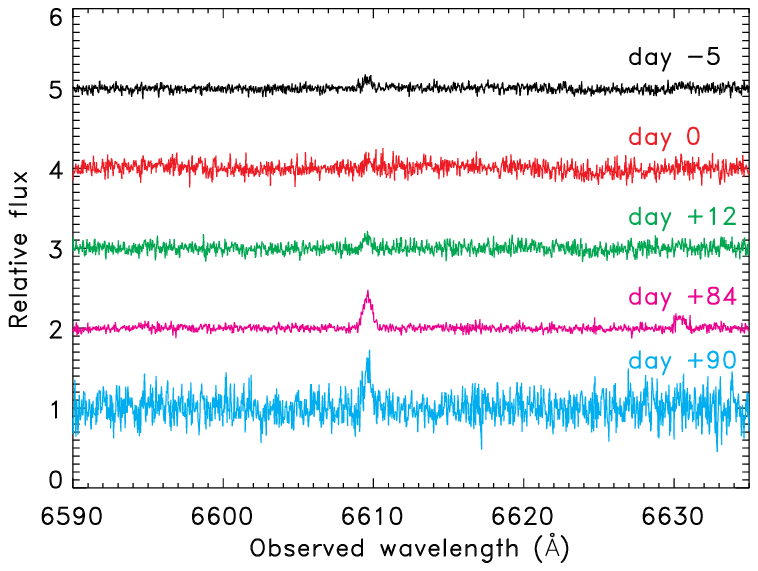}
\caption{High-resolution spectra of SN~2007le around the redshifted
  \ha\ and [\ion{N}{2}] lines.  From top to bottom, the spectra are
  from days~$-5$, 0, $+12$, $+84$, and $+90$ (the day~$+10$ spectrum
  is excluded because local sky subtraction cannot be done on a fiber
  spectrum).  A constant offset has been added to the top four spectra
  for clarity, and the days~$-5$, 0, and $+12$ spectra have been
  multiplied by a factor of 3 before adding the offset to make the
  emission line visible on this display scale.  The emission line
  appears to be stronger at late times primarily because the supernova
  continuum on top of which it is superimposed has faded by more than
  an order of magnitude.}
\label{ha_1dspec}
\end{figure}

We detect a narrow \ha\ emission line at the position of the supernova
at all epochs of our observations.  The HIRES spectra were obtained
with a slit 7\arcsec\ long and either 0.86\arcsec\ or
1.15\arcsec\ wide, and faint \ha\ emission from the host galaxy is
visible all the way along the slit.  However, careful sky subtraction
reveals that there is also additional emission that is spatially
unresolved and coincident with the position of the supernova (see
Figures \ref{ha_2dspec} and \ref{ha_1dspec}).  This line is also at a
slightly different velocity than the extended host-galaxy emission.

From the HIRES data, we measure the EW, velocity, and line width of
the \ha\ emission at the position of the supernova; the results are
listed in Table \ref{ha_ew}.  At $v = 2139$~\kms, the \ha\ emission is
offset by 16~\kms\ from the velocity of the variable sodium line (and
has a higher velocity than any of the main sodium components), but is
similar to the host-galaxy rotation velocity at the radius of the
supernova measured from a long-slit spectrum by \citet{afanasyev92}.
We then use the observed light curve and a comprehensive set of
low-resolution spectra of SN~2007le from the CfA Supernova Archive to
derive absolute flux calibrations for the spectra.  In order to remove
contaminating host-galaxy light (since the CfA spectra were obtained
with a wide slit), the spectra are first scaled to the observed
$V$-band magnitudes of the supernova and then warped to match the $B$
magnitudes as well.  From these flux-calibrated spectra, we determine
the SN continuum level at the position of \ha\ (6610~\AA), scaling the
spectra up and down as necessary to account for the change in the
supernova's $R$ magnitude between the time of observation for each
high-resolution spectrum and the nearest low-resolution spectrum (the
time differences are less than two days for days~$-5$ and $+12$, six
days for day~0, and five and eleven days for days~$+84$ and $+90$,
respectively).  We then convert the high-resolution EWs to fluxes in
absolute units with the measured continuum levels.

In principle, the \ha\ emission could originate either from an
\hii\ region projected within 140~pc (corresponding to the
$\approx$1\arcsec\ slit width) of the explosion or from the supernova
itself.  In the case of an \hii\ region, one would expect other lines
such as [\ion{O}{3}]~$\lambda\lambda$4959, 5007,
[\ion{N}{2}]~$\lambda\lambda$6548, 6583, and
[\ion{S}{2}]~$\lambda\lambda$6717, 6731 to be visible depending on the
S/N.  In the day~$+84$ spectrum, which is the best for this purpose
because of its relatively high S/N at red wavelengths and the faint
magnitude of the SN at late times, we indeed detect H$\beta$ and both
lines of the [\ion{N}{2}] and [\ion{S}{2}] doublets.  [\ion{O}{3}],
though, which is generally the brightest of the forbidden metal lines
in \hii\ regions (and comparable to or brighter than H$\beta$),
remains undetected at all epochs.  [\ion{O}{3}] fluxes more than an
order of magnitude weaker than \ha\ are uncommon in \hii\ regions
\citep*[e.g.,][]{kennicutt03,magrini07}.

On the other hand, if the \ha\ emission is related to the supernova,
changes in the \ha\ flux with time would seem likely (although since
such emission has never been detected, its exact characteristics are
unknown).  From the day $-5$ spectrum through the day~$+12$ spectrum
(spanning about 2.5 weeks), we detect no variability in the flux.  In
the day~$+84$ spectrum more than two months later, however, the
\ha\ flux is $\sim40$\% higher than at day~$+12$.  If the comparison
is made only between days~$+12$ and $+84$, the significance of this
change is just 2~$\sigma$, but if we combine the measurements from
days~$-5$, 0, and $+12$ together then the day~$+84$ increase is
significant at the 4~$\sigma$ level.  The low-S/N day~$+90$ spectrum
is consistent with both the early time and day~$+84$ measurements.
Taken as a whole, we consider these results to be a tentative
indication of \ha\ emission from SN~2007le.  More conclusive evidence
of the nature of the \ha\ emission awaits late-time imaging and
spectroscopy to determine whether or not there is an \hii\ region
coincident with the position of the supernova.

If the \ha\ emission is associated with the SN, its luminosity is
related to the amount of gas present.  The observed \ha\ emission has
a luminosity of $1.6 \times 10^{35}$~ergs~s$^{-1}$ (for the average
\ha\ flux of $1.5 \times 10^{-16}$~ergs~cm$^{-2}$~s$^{-1}$).  If we
follow the calculations of \citet{leonard07}, such a luminosity
corresponds to $\sim0.025$~M$_{\odot}$ of solar-abundance material in
the circumstellar environment, in reasonable agreement with
theoretical predictions \citep*{marietta00,mattila05,pakmor08}.
However, those computations were specifically aimed at modeling
\ha\ emission at very late times (day~$+380$) and may not apply to the
much earlier epochs at which we observed SN~2007le.  Using the simple
model presented by \citet{patat} as an alternative suggests a hydrogen
mass of $\sim0.001$~M$_{\odot}$ for the observed \ha\ luminosity.  The
presence of hydrogen in the CSM has also been indirectly inferred from
the broadening of absorption features in the earliest spectra of
SNe~Ia \citep{mazzali05a}.  In this scenario, $\sim 0.005$~M$_{\odot}$
of material with a solar composition is necessary to provide the
electron density that is required to trap photons and cause absorption
in lines such as Ca~H \& K or \ion{Si}{2}~$\lambda$6355
\citep{mazzali05b,tanaka08}.

\section{DISCUSSION}
\label{discussion}

For SN~2006X, \citet{patat} suggested a model in which the progenitor
system consisted of a white dwarf and a red giant, with either the
stellar wind from the red giant or repeated nova eruptions blowing off
several possibly asymmetric shells of material before the explosion.
These circumstellar shells were proposed to be the location of the
material responsible for the variable Na~D absorption.  Given
timescales of the order of decades for the mass loss and the observed
velocities of the variable lines, the absorbing material would be
$\sim10^{16}$~cm away from the star at the time of explosion.  This
scenario envisions that the \ion{Na}{1} atoms (with an ionization
potential of 5.1~eV) are ionized by the UV radiation from the
supernova, and then slowly recombine over the following weeks to
produce strengthening Na~D absorption lines.  The absence of
variations in the Ca~H \& K lines is then attributed to the much
higher ionization potential of \ion{Ca}{2} (11.9~eV), which prevents
most of the Ca ions from being affected by the SN radiation field.

However, \citet{chugai} has challenged this interpretation, arguing
that the physical conditions expected in a red giant wind are not
compatible with strong Na~D absorption.\footnote{Note, however, that
  the underprediction of \ion{Na}{1} is a known issue in ionization
  models \citep[e.g.,][]{mazzali97}.  The reason for this problem is
  not understood, but it could provide an explanation for the observed
  EW ratio between \ion{Na}{1} and \ion{Ca}{2} exceeding the expected
  one by a large factor (see \S \ref{cloudy}). }  In light of these
calculations, plus the confirmation from SN~2007le and SN~1999cl
\citep{blondin08} that the variable Na absorption phenomenon is not
unique to SN~2006X, we now reconsider the nature of the absorbing
material.

\subsection{Photoionization and Recombination Timescales}
\label{ion_recomb}

If the changes in the Na~D absorption lines are a result of
recombination, then the spectra tell us directly that the
recombination timescale is $\sim10$~days (e.g.,
Fig.~\ref{highresspectra}).  Since

\begin{equation}
\tau_{\rm recomb} = \frac{1}{\alpha n_{e}},
\label{eq_recomb}
\end{equation}

\noindent
where $\alpha$, the \ion{Na}{1} recombination coefficient, is $1.43
\times 10^{-13}$~cm$^{3}$~s$^{-1}$ at $T = 10^{4}$~K \citep[][see \S
  \ref{cloudy} for a justification of this temperature
  range]{badnell06}, we find that the electron density must be $n_{e}
\approx 8 \times 10^{6}$~cm$^{-3}$.  (For $T = 2 \times 10^{3}$~K,
$\alpha = 6.18 \times 10^{-13}$~cm$^{3}$~s$^{-1}$, and the required
electron density will be lower by a corresponding factor.)  We note
that such a high electron density necessitates that hydrogen must be
partially or mostly ionized; no other plausible source could provide
so many electrons (as mentioned in \S~\ref{ha}, \citealt{mazzali05a}
suggest that electrons provided by hydrogen can also explain the
origin of high-velocity features, as were present in SN~2007le, at
early times).  Even with full ionization, though, the physical density
required in the absorbing material is quite high.

In order for a significant fraction of neutral sodium to be present,
the photoionization rate should be comparable to or slower than the
recombination rate, or alternatively, the recombination time should be
shorter than the photoionization time.  Following \citet{murray07}, we
write

\begin{equation}
\tau_{ion} = \frac{4 \pi r^{2} \langle h\nu \rangle}{a_{\rm Na~I} L_{\rm UV}},
\label{ion_time}
\end{equation}

\noindent
where $r$ is the distance between the SN and the absorbing material,
$a_{\rm Na~I} \approx 10^{-19}$~cm$^{2}$ is the photoionization cross
section \citep{verner96}, $L_{\rm UV}$ is the UV luminosity of the
supernova, and $\langle h\nu \rangle$ corresponds to the photon energy
required to ionize a \ion{Na}{1} atom.  We do not have UV photometry
of SN~2007le, but to the degree that UV light curves of SNe~Ia are
fairly homogeneous \citep{brown08}, it is reasonable to substitute
another similar SN.  We therefore use the {\it Swift} light curve of the
well-observed normal SN~Ia 2007af from \citeauthor{brown08} to
estimate the UV luminosity.  SN~2007af reached a peak Vega magnitude
of $\sim 16.2$ in the {\it Swift} $UVW2$ filter a few days before the
$B$-band maximum.  The $UVW2$ filter has a central wavelength of
1928~\AA, a full width at half-maximum intensity (FWHM) of 
657~\AA, and a photometric zeropoint of $17.35
\pm 0.03$~mag \citep{swiftcal}.  We use the distance modulus of
SN~2007af, 32.06~mag \citep{simon07}, to deduce an absolute magnitude
of $M_{UVW2} = -15.86$, which suggests an apparent peak magnitude for
SN~2007le of 16.49.  With the zeropoint of the $UVW2$ filter, this
magnitude translates to a {\it Swift} count rate of 2.2~counts~s$^{-1}$.
\citet{swiftcal} provide a conversion between the count rate and the
flux density that depends only slightly on the source spectrum
($\sim6.1 \times 10^{-16}$~ergs~cm$^{-2}$~s$^{-1}$~\AA$^{-1}$), so
integrating over the filter bandwidth yields a flux of $8.8 \times
10^{-13}$~ergs~cm$^{-2}$~s.  We therefore estimate a $UVW2$ luminosity
for SN~2007le of $9 \times 10^{40}$~ergs~s$^{-1}$.  Note that this
bandpass corresponds reasonably closely to the wavelengths that can
ionize \ion{Na}{1} atoms most efficiently ($\lambda < 2412$~\AA).
Using Equation \ref{ion_time}, we can now calculate that the
ionization timescale is

\begin{equation}
\tau_{\rm ion} = 1.1 \left( \frac{r}{10^{16}~\mbox{cm}} \right)^{2}~\mbox{s}.
\label{eq_ion}
\end{equation}

\noindent
For $r \lta 3$~pc, then, $\tau_{\rm ion} \ll \tau_{\rm recomb}$, and this
estimate suggests that the Na should be essentially fully ionized
around maximum light.

Although the UV luminosity of the SN declines with time, the {\it Swift}
data show that the decline is rather slow: $\sim$1~mag in 10~days
\citep{brown08}.  Even several weeks after peak, when the
high-resolution spectroscopy demonstrates that significant changes in
the Na~D line profile have already occurred, Equations \ref{eq_recomb}
and \ref{eq_ion} indicate that recombination (at the assumed
densities) should only be occurring at relatively large radii, calling
into question the hypothesis that \emph{circumstellar} material is
responsible for the varying Na absorption.  These simple calculations
are in qualitative agreement with the results of \citet{chugai} and
indicate that more sophisticated modeling is required.

\subsection{Photoionization Modeling}
\label{cloudy}

Because the simple ionization and recombination timescale arguments do
not provide obvious answers as to why the \ion{Na}{1} column density
is changing, we now consider more detailed photoionization models of
the SN environment.  Starting with a synthetic UV spectrum for a Type
Ia supernova from \citet{nugent95} as the ionizing source, we carry
out calculations with the photoionization code Cloudy
\citep[][ver. 07.02]{ferland98}.  We use the UV spectra of SN~2001eh
and SN~2001ep \citep{sauer08} to normalize the total number of UV--IR
photons ($1500 - 12000$~\AA) in the model spectrum.\footnote{The
  normalization $N_{\rm SN}$ is chosen such that $4 \pi d_{\rm SN}^2
  \int \lambda J^{\rm SN}_\lambda d\lambda = N_{\rm SN} \int \lambda
  J^{\rm model}_\lambda d\lambda$, where $J^{\rm model}_\lambda$ is
  the model spectrum, and $J^{\rm SN}_\lambda$ and $d_{\rm SN}$ are
  the SN spectrum and distance.  The normalization factors for
  SN~2001ep and SN~2001eh agree to within $30\%$, so we use $N_{\rm
    SN} = 0.5(N_{\rm ep} + N_{\rm eh})$.}  The model spectrum is only
available for the time of maximum brightness, so we assume that the
light curve at all UV wavelengths follows the SN~2007af $UVW2$ light
curve from \citet{brown08}.  (We have to combine observations of
multiple SNe here because there are few nearby type Ia supernovae with
both good UV light curves and UV spectroscopy.)  This assumption means
that the calculations are not explicitly time dependent, but we do not
expect non-equilibrium effects to produce qualitative changes in the
results discussed below.  We also note that, while at longer UV
wavelengths the model can be compared to observed spectra for
validation, at the critical far-UV wavelengths that ionize hydrogen no
such comparison with real data is possible.  \citet*{borkowski09} have
recently carried out very similar calculations to investigate the
possibility of detecting CSM interactions in spectra obtained before
maximum light.

\begin{figure*}[t!]
\epsscale{1.24}
\plotone{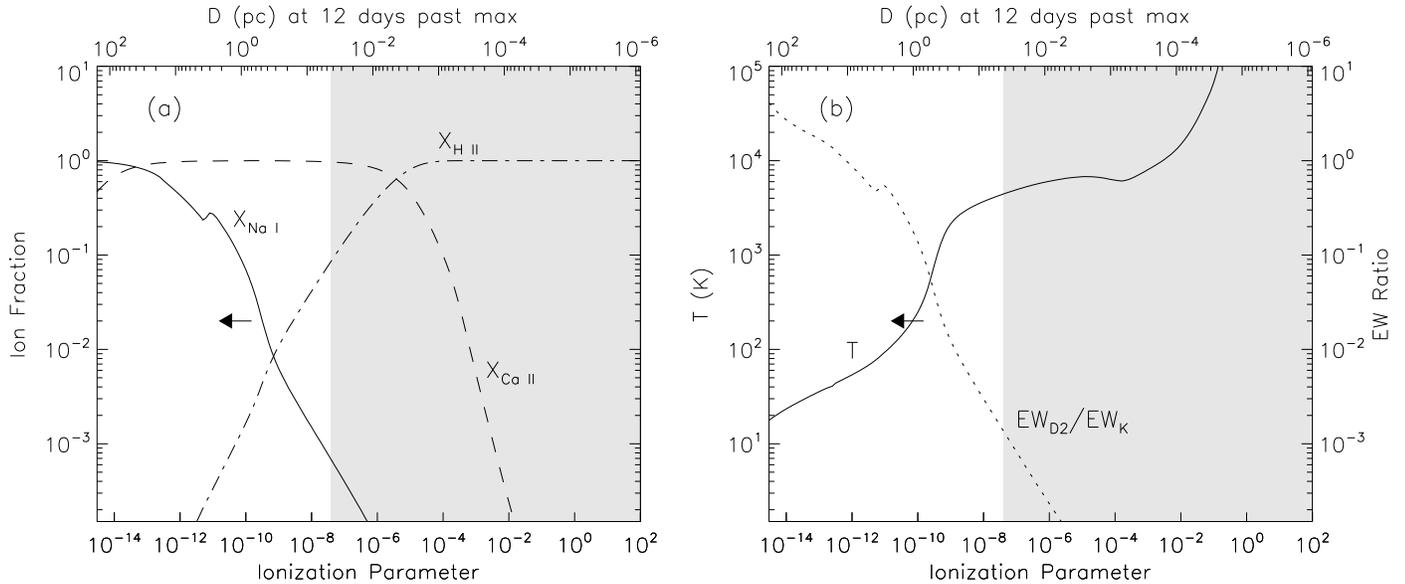}
\caption{Results of photoionization calculations performed with Cloudy
  for a total density of $n=2.4 \times 10^{7}$~cm$^{-3}$ 12 days after
  maximum brightness.  (\emph{a}) Ion fractions for H, Na, and Ca as a
  function of ionization parameter (or distance along the top axis).
  The solid line shows the \ion{Na}{1} fraction, the dashed line shows
  the \ion{Ca}{2} fraction, and the dot-dashed line shows the
  \ion{H}{2} fraction.  The black arrow illustrates the magnitude of
  the change in ionization parameter from maximum light to 30 days
  later, and the shaded gray region on the right side of the plot
  represents the range of ionization parameters for which the dust
  grain equilibrium temperature is larger than the dust sublimation
  temperature, so that dust grains should not survive (and thus Ca
  atoms should all be in the gas phase).  (\emph{b}) Temperature and
  equivalent width ratio as a function of ionization parameter.  The
  dotted line represents the ratio of the equivalent width of the
  \ion{Na}{1}~D$_{2}$ line to that of the \ion{Ca}{2}~K line (axis
  scale on the right), and the solid line shows the temperature (axis
  scale on the left).  The Cloudy model demonstrates that variable Na
  absorption, constant Ca absorption, and highly depleted Ca
  abundances can be obtained for distances larger than $\sim0.1$~pc
  away from the site of the explosion.}
\label{cloudymodel}
\end{figure*}

Guided by the hypothesis that recombination is responsible for the
varying \ion{Na}{1} column density, a few iterations of the model
demonstrate that very large hydrogen densities are needed to produce
significant Na recombination (for a solar abundance ratio).  We
therefore assume a hydrogen density of $n=2.4 \times 10^{7}$~cm$^{-3}$
(somewhat higher than the electron density inferred in
\S~\ref{ion_recomb}).  We then calculate the \ion{Na}{1}, \ion{Ca}{2},
and \ion{H}{2} fractions as a function of time and distance away from
the SN.  The results are illustrated in Figure \ref{cloudymodel}.  For
distances greater than $\sim0.1$~pc the fraction of \ion{Ca}{2} ions
is essentially constant with time, while the \ion{Na}{1} fraction
increases by approximately an order of magnitude over a time span of
30 days.  Closer to the SN, the Ca becomes more highly ionized (and
the \ion{Ca}{2} fraction varies with time) and the amount of neutral
Na is negligible, supporting the idea that the source of the
absorption must be located at distances larger than $10^{16}$~cm.

Over the range of parameter space where these calculations indicate
that the \ion{Na}{1} column density will vary and the \ion{Ca}{2}
column density will not, the EW of the Na~D$_{2}$ line is predicted to
be less than or equal to the EW of the Ca~K line.  At distances of
less than 1~pc, the Ca absorption should be $\gta100$ times as strong
as the Na absorption.  Yet, we observe that EW$_{{\rm Na~D}}$ is
\emph{larger} than EW$_{{\rm Ca~H\&K}}$ at all epochs (see Tables
\ref{na_ew} and \ref{ca_ew} and Fig. \ref{highresspectra_ca}).  If the
model described above is correct, we therefore require that nearly all
of the Ca atoms in the CSM are locked up in dust grains, as is also
typical in dense interstellar clouds
\citep*[e.g.,][]{spitzer54,howard63,herbig68,morton75}.  In the
immediate vicinity of a supernova, the destruction of dust grains is
obviously a possibility.  However, given sublimation temperatures of
$\sim1500$~K for silicate and carbonaceous grains \citep{ds79},
calculations of grain temperatures with Cloudy using the dust grain
model of \citet{vanhoof04} indicate that dust can survive at distances
greater than 0.05~pc (unshaded region in Fig. \ref{cloudymodel}) over
the relevant time period.

\subsection{Physical Conditions in the Absorbing Gas}

The results presented in \S~\ref{cloudy} point to distances of
$\sim0.1$~pc for the absorbing material.  Figure \ref{cloudymodel}
demonstrates that the EW of the \ion{Ca}{2} absorption would remain
constant for distances as small as $\sim0.02$~pc, but the fraction of
neutral sodium so close to the supernova is negligible.  Distances
greater than 1~pc are also allowed by the photoionization modeling,
but then the ionization fraction is very low (as a result of the weak
UV field), requiring extremely high physical densities to produce
large numbers of electrons.  Of course, at such distances the galactic
UV field may contribute significantly to the photoionization rate,
rendering our calculations no longer applicable.  At $d = 0.1$~pc, the
gas temperature is $\sim5000$~K, and densities of $n_{{\rm H}} \approx
2.4 \times 10^{7}$~cm$^{-3}$ are required to produce significant Na
recombination.  With such a high density, substantial amounts of dust
could be present, and strong depletion of Ca atoms onto dust grains
would explain the large observed ratio of \ion{Na}{1} to \ion{Ca}{2}.
Given the derived density and a solar Na abundance of 12 + log(Na/H)$
= 6.17$ \citep*{ags05}, the total density of Na atoms is $n_{{\rm Na}}
= 35$~cm$^{-3}$.  The \ion{Na}{1} fraction under these conditions is
$\sim5 \times 10^{-4}$, yielding $n_{{\rm Na~I}} = 1.8 \times
10^{-2}$~cm$^{-3}$.  In order to obtain a column density of $\sim2.5
\times 10^{12}$~cm$^{-2}$, as derived in \S~\ref{conditions}, the path
length through the absorbing cloud must be $\sim1.4 \times
10^{14}$~cm.  If such a clump of material were roughly spherical its
mass could be quite small ($\sim2 \times 10^{-7}$~M$_{\odot}$), but a
clump of radius $10^{14}$~cm at a distance of $10^{17}$~cm would only
occult a small fraction of the SN photosphere (radius
$\sim10^{15}$~cm) after maximum light and thus could not be
responsible for an absorption line as deep as that observed in
Figure~\ref{highresspectra}.  In order to produce a covering fraction
closer to unity, there must either be large numbers of small clumps,
or perhaps a larger sheet of material with perpendicular dimensions of
$\sim10^{17}$~cm (the rather high densities involved probably rule out
the full spherical shell model proposed by \citealt{patat}).  The
total CSM mass along the line of sight (presumably more CSM is present
in other directions) would then be in the range of $10^{-5} -
10^{-2}$~M$_{\odot}$ depending on the geometry.

Is it plausible for such a cloud (or clouds) to exist in the SN
progenitor system?  The measured velocity of the variable absorbing
component is 2123~\kms, while the local host galaxy ISM velocity near
the position of the SN is $\approx2130$~\kms.  Assuming that the
velocity measured for the local ISM is representative of the
progenitor system, then the variable absorption component has a
line-of-sight velocity of $\sim10$~\kms\ relative to the progenitor
itself.  At a constant velocity of 10~\kms, material starting near the
center of the progenitor system would take $\sim10^{4}$~yr to reach a
radius of 0.1~pc, but factoring in the gravitational deceleration the
actual travel time would be $\sim3000$~yr if the material originated
in the red giant wind or $\sim10$~yr if it was blown off from the
white dwarf (which requires a much higher initial velocity).
Regardless, this timescale is certainly far shorter than the age of
the progenitor system or its evolutionary timescale, so
\emph{producing} CSM at such distances does not appear to be
problematic.  The \emph{survival} of dense clumps similar to what we
are proposing here, though, may be an issue: the thermal pressure in
such a cloud is quite large, suggesting that the cloud should expand
and dissipate on timescales of decades unless the average CSM pressure
is also very high.  We do not currently have enough information to
determine whether such clumps of material could form from the wind of
the companion alone, the interaction of the wind with surrounding
material, or a nova eruption on the surface of the white dwarf.

\section{SUMMARY AND CONCLUSIONS}
\label{conclusions}

We have presented a series of high-resolution spectra of the nearby
Type Ia supernova SN~2007le, a normal SN~Ia in the subgroup that
exhibits broad lines and high velocity gradients, similar to SN~2002bo
and SN~2006X.  The spectra reveal a complex set of absorption systems
in the Na~D and Ca~H \& K lines.  Our observations demonstrate that
one of the Na absorption components changes systematically with time,
increasing its equivalent width by 85~m\AA\ from maximum light to
three months later.  SN~2007le is the third SN~Ia in which this effect
has been seen, after SN~2006X \citep{patat} and SN~1999cl
\citep{blondin08}.  As in SN~2006X, we detect no variations in the
\ion{Ca}{2} absorption column over the course of our observations.

In the initial study of SN~2006X, \citet{patat} modeled the varying
absorption as occurring in several shells of circumstellar gas
resulting either from the stellar wind of the companion to the
supernova progenitor (presumed to be a red giant) or recurrent nova
eruptions.  Subsequently, \citet{chugai} argued that the red giant
wind scenario could only reproduce the observed behavior with an
unrealistically high wind density.  \citet{chugai} did note that a red
giant wind plus a dense clump or shell of material might be compatible
with the observed \ion{Na}{1} and \ion{Ca}{2} optical depths, but
concluded that this solution was unlikely because of the difficulty of
creating a sufficiently dense shell.  Moreover, both SN~2006X and
SN~1999cl were extremely heavily reddened, suggesting the possibility
of a connection between variable absorption and properties of the
interstellar medium along the line of sight.  The much lower
extinction toward SN~2007le demonstrates that variable absorption also
occurs in supernovae that are not seen through an unusually dense,
dusty ISM.  Conversely, it is now also clear that some highly obscured
objects such as SN~2003cg \citep{blondin08} and SN~2008fp (Patat et
al., in preparation) do not exhibit variable absorption.  We therefore
carried out further photoionization calculations to investigate
whether circumstellar material could explain the varying Na and
constant Ca absorption profiles.  Using a synthetic UV spectrum for a
SN~Ia, we found that a small ($10^{14}$~cm), dense
($10^{7}$~cm$^{-3}$) clump or clumps of material located $\sim0.1$~pc
away from the explosion can provide the necessary column density and
ionization changes with time to account for the observations.  In
order to reproduce the observed \ion{Na}{1} to \ion{Ca}{2} EW ratio,
this model requires a very large fraction of the Ca atoms to be
depleted onto dust grains.  The most likely origin for gas so close to
the supernova is that it was produced in the progenitor system.  If
this explanation is correct, the results indicate a single-degenerate
progenitor system for SN~2007le, in accord with previous suggestions
for SN~2006X.

Through the work of \citet{blondin08} and our own high-resolution
spectroscopy of additional SNe~Ia (to be described in future papers),
it is clear that the fraction of SN~Ia events that show the variable
Na signature is relatively small (as predicted by
\citealt{patat,patat07b} for the recurrent nova scenario).  In light
of this rarity, it is noteworthy that all three of the SNe displaying
variable absorption share common features in their early-time and
maximum-light spectra, classifying them in the broad line group
according to \citet{branch09} and the high velocity gradient group of
\citet[][although the HVG classification of SN~2007le is less
  firm]{benetti05}.  With these groups comprising $\sim30$\% of
SNe~Ia, the odds of three SNe exhibiting both variable absorption and
broad lines/high velocity gradients by chance (i.e., if there is no
physical connection between the two) is low.  Our observations of
SN~2007le therefore support earlier speculations of a relationship
between high velocity gradients and variable Na absorption.

These results still leave open the three explanations for variable
absorption proposed in the literature: (1) SNe with high velocity
gradients and variable absorption occur in a distinct progenitor
system or with a different explosion mechanism than other SNe~Ia; (2)
SNe with high velocity gradients and variable absorption occur in the
same progenitor systems as other SNe~Ia, but with a peculiar geometry
that yields higher observed velocities and makes variable absorption
more likely; or (3) variable absorption is simply a line-of-sight
effect that is independent of the properties of the supernova itself.
If the variable absorption originates in the ISM, then it is much more
likely to be observed for objects that are seen through large ISM
columns.  The low extinction of SN~2007le, suggesting that it is not
obscured by a substantial amount of ISM material, renders the third
possibility less likely.  Further high-resolution spectroscopy of
BL/HVG SNe will be needed to continue investigating the first two
scenarios and reveal whether high velocity gradients are a sufficient
condition for variable Na absorption or merely a necessary one.  It is
also worth noting that in all three SNe with variable absorption, the
absorption lines are systematically blueshifted with respect to the
local ISM velocity, further pointing to a connection between the
absorbing material and the supernovae themselves.

In addition to variable Na absorption, we also detect \ha\ emission in
the spectrum of SN~2007le, possibly originating from the supernova
itself.  The narrow \ha\ emission line (FWHM~$ = 25$~\kms) is
spatially unresolved, coincident with the position of the supernova,
and at a slightly different velocity than the surrounding diffuse
interstellar emission.  The amount of hydrogen that would need to be
present in the circumstellar environment to produce the observed
emission luminosity is $\sim10^{-2} - 10^{-3}$~M$_{\odot}$.  This feature is
present at all high-resolution spectroscopic epochs, but as a result
of its small equivalent width is not visible in any of the
low-resolution spectra.  The emission-line flux is constant within the
uncertainties through 12 days after maximum light, but possibly
strengthens in the observations obtained several months later.  In the
best of our spectra, H$\beta$, [\ion{N}{2}], and [\ion{S}{2}] lines
are also detected, but interestingly [\ion{O}{3}] is not, suggesting a
very low excitation for the ionized gas.  If the increase in the
\ha\ flux with time is real, then the emitting gas must be associated
with the supernova, but further observations at late times after the
SN has faded will be needed to confirm or reject this possibility.

\acknowledgements{The authors wish to acknowledge the very significant
  cultural role and reverence that the summit of Mauna Kea has always
  had within the indigenous Hawaiian community.  We are most fortunate
  to have the opportunity to conduct observations from this mountain.
  The HET is named in honor of its principal benefactors, William
  P. Hobby and Robert E. Eberly.  We thank Xavier Prochaska for his
  work on developing the HIRES data reduction pipeline and answering
  all of our extensive questions about it.  We also thank the
  anonymous referee for constructive suggestions and acknowledge
  helpful conversations with Chris Burns, Mark Phillips, Doug Leonard,
  and Juna Kollmeier.  Some calculations described in this paper were
  performed with version 07.02 of Cloudy, last described by
  \citet{ferland98}.  This research has made use of NASA's
  Astrophysics Data System Bibliographic Services and the NASA/IPAC
  Extragalactic Database (NED), which is operated by the Jet
  Propulsion Laboratory, California Institute of Technology, under
  contract with the National Aeronautics and Space Administration.

  J.D.S. acknowledges the support of a Millikan Fellowship provided by
  Caltech and a Vera Rubin Fellowship from the Carnegie Institution of
  Washington.  A.G. acknowledges support by the Israeli Science
  Foundation; an EU Seventh Framework Programme Marie Curie IRG
  fellowship; the Ministry of Science, Culture \& Sport, Israel and
  the Ministry of Research, France; and the Benoziyo Center for
  Astrophysics, UK-Weizmann fund, a research grant from the Peter and
  Patricia Gruber Awards, and the William Z. and Eda Bess Novick New
  Scientists Fund at the Weizmann Institute.  R.Q. and J.C.W.  are
  supported in part by NSF grant AST--0707769.  A.V.F.'s supernova
  group at U.C. Berkeley is supported by NSF grant AST--0607485, US
  Department of Energy grant DE-FG02-08ER41563, and the TABASGO
  Foundation. KAIT and its ongoing operation were made possible by
  donations from Sun Microsystems, Inc., the Hewlett-Packard Company,
  AutoScope Corporation, Lick Observatory, the NSF, the University of
  California, the Sylvia \& Jim Katzman Foundation, and the TABASGO
  Foundation.  Supernova research at the Harvard College Observatory
  is supported in part by the NSF through AST--0606772.  }

{\it Facilities:} Keck:I (HIRES)

\clearpage
\LongTables
\begin{landscape}
\begin{deluxetable}{llcclllll}
\tablewidth{0pt}
\tabletypesize{\scriptsize}
\tablecolumns{9}
\tablecaption{Emission-Line Widths and Fluxes}
\tablehead{
\colhead{Epoch} & \colhead{\ha\ EW} & 
\colhead{\ha\ velocity} & \colhead{\ha\ line width} & 
\colhead{\ha\ flux} & \colhead{H$\beta$ flux} &
\colhead{[\ion{O}{3}] flux} & \colhead{[\ion{N}{2}] flux\tablenotemark{a}} &
\colhead{[\ion{S}{2}] flux\tablenotemark{b}} \\
\colhead{} & \colhead{[m\AA]} & 
\colhead{[\kms]} & \colhead{[\kms]} & 
\colhead{[ergs cm$^{-2}$ s$^{-1}$]} & 
\colhead{[ergs cm$^{-2}$ s$^{-1}$]} & 
\colhead{[ergs cm$^{-2}$ s$^{-1}$]} & 
\colhead{[ergs cm$^{-2}$ s$^{-1}$]} & 
\colhead{[ergs cm$^{-2}$ s$^{-1}$]} }
\startdata 
day $-5$ & $22 \pm 3$   & $2137.8 \pm 1.1$ & $25.9 \pm 4.5$ & $(1.5 \pm 0.2) \times 10^{-16}$ & ...\tablenotemark{c} &  $<5.8 \times 10^{-17}$ & ... & ... \\
day 0 & $14 \pm 4$   & $2140.1 \pm 1.7$ & $17.9 \pm 4.0$  & $(1.2 \pm 0.3) \times 10^{-16}$ & ... & ... & ... & ... \\
day $+12$ & $28 \pm 5$   & $2136.4 \pm 1.4$ & $25.4 \pm 3.3$ & $(1.5 \pm 0.3) \times 10^{-16}$ & ... & ... & ... & ... \\
day $+84$ & $288 \pm 9$  & $2139.3 \pm 0.3$ & $30.8 \pm 0.7$ & $(2.1 \pm 0.1) \times 10^{-16}$ & $3.9 \pm 0.7 \times 10^{-17}$  & $<1.5 \times 10^{-17}$ & $(6.0 \pm 0.7) \times 10^{-17}$ & $(2.7 \pm 0.5) \times 10^{-17}$ \\
day $+90$ & $301 \pm 41$ & $2139.5 \pm 1.0$ & $24.5 \pm 2.6$ & $(1.8 \pm 0.3) \times 10^{-16}$ & ... & ... & ... & ... \\

\enddata 
\tablenotetext{a}{The listed value is for the stronger
  [\ion{N}{2}]~$\lambda6583$ line.}
\tablenotetext{b}{The listed value is for the stronger
  [\ion{S}{2}]~$\lambda6717$ line.}
\tablenotetext{c}{We have listed upper limits only where the data
  provide useful constraints; [\ion{O}{3}] upper limits are at
  4~$\sigma$ significance}.
\label{ha_ew}
\end{deluxetable}
\clearpage
\end{landscape}


\begin{thebibliography}{}

\bibitem[Afanasyev et al.(1992)]{afanasyev92} Afanasyev, V.~L.,
  Burenkov, A.~N., Zasov, A.~V., \& Silchenko, O.~K.\ 1992, \azh, 69,
  19

\bibitem[Aldering(2005)]{aldering05} Aldering, G.\ 2005, New Astronomy
  Review, 49, 346

\bibitem[Aldering et al.(2006)]{aldering06} Aldering, G., et
  al.\ 2006, \apj, 650, 510

\bibitem[Asplund et al.(2005){Asplund, Grevesse, \& Sauval}]{ags05}
  Asplund, M., Grevesse, N., \& Sauval, A.~J.\ 2005, in ASP
  Conf. Ser. 336, Cosmic Abundances as Records of Stellar Evolution
  and Nucleosynthesis, ed. T.~G.~Barnes~III \& F.~N.~Bash (San
  Francisco, ASP), 25

\bibitem[Astier et al.(2006)]{astier06} Astier, P., et al.\ 2006,
  \aap, 447, 31

\bibitem[Badenes et al.(2006)]{badenes06} Badenes, C., Borkowski,
  K.~J., Hughes, J.~P., Hwang, U., \& Bravo, E.\ 2006, \apj, 645, 1373

\bibitem[Badnell(2006)]{badnell06} Badnell, N.~R.\ 2006, \apjs, 167,
  334

\bibitem[Benetti et al.(2005)]{benetti05} Benetti, S., et al.\ 2005,
  \apj, 623, 1011

\bibitem[Benetti et al.(2006)]{benetti06} Benetti, S., Cappellaro, E.,
  Turatto, M., Taubenberger, S., Harutyunyan, A., \& Valenti,
  S.\ 2006, \apjl, 653, L129

\bibitem[Benetti et al.(2004)]{benetti04} Benetti, S., et al.\ 2004,
  \mnras, 348, 261

\bibitem[Blondin et al.(2009)]{blondin08} Blondin, S., Prieto, J.~L.,
  Patat, F., Challis, P., Hicken, M., Kirshner, R.~P., Matheson, T.,
  \& Modjaz, M.\ 2009, \apj, 693, 207

\bibitem[Blondin \& Tonry(2007)]{bt07} Blondin, S., \& Tonry,
  J.~L.\ 2007, \apj, 666, 1024

\bibitem[Bode et al.(2007)]{bode07} Bode, M.~F., Harman, D.~J.,
  O'Brien, T.~J., Bond, H.~E., Starrfield, S., Darnley, M.~J., Evans,
  A., \& Eyres, S.~P.~S.\ 2007, \apjl, 665, L63

\bibitem[Borkowski et al.(2009){Borkowski, Blondin, \&
    Reynolds}]{borkowski09} Borkowski, K.~J., Blondin, J.~M., \&
  Reynolds, S.~P.\ 2009, \apjl, 699, L64

\bibitem[Branch et al.(2009){Branch, Dang, \& Baron}]{branch09}
  Branch, D., Dang, L.~C., \& Baron, E.\ 2009, \pasp, 121, 238

\bibitem[Branch et al.(2006)]{branch06} Branch, D., et al.\ 2006,
  \pasp, 118, 560

\bibitem[Brown et al.(2009)]{brown08} Brown, P.~J., et al.\ 2009, \aj,
  137, 4517

\bibitem[Chandrasekhar(1931)]{chandra} Chandrasekhar, S.\ 1931,
  \mnras, 91, 456

\bibitem[Chugai(2008)]{chugai} Chugai, N.~N.\ 2008, Astronomy Letters,
  34, 389

\bibitem[Cox \& Patat(2008)]{cp08} Cox, N.~L.~J., \& Patat, F.\ 2008,
  \aap, 485, L9

\bibitem[Crotts \& Yourdon(2008)]{crotts08} Crotts, A.~P.~S., \&
  Yourdon, D.\ 2008, \apj, 689, 1186

\bibitem[Cumming et al.(1996)]{cumming96} Cumming, R.~J., Lundqvist,
  P., Smith, L.~J., Pettini, M., \& King, D.~L.\ 1996, \mnras, 283,
  1355

\bibitem[de Vaucouleurs \& Corwin(1985)]{dvc85} de Vaucouleurs, G., \&
  Corwin, H.~G., Jr.\ 1985, \apj, 295, 287

\bibitem[Draine \& Salpeter(1979)]{ds79} Draine, B.~T., \& Salpeter,
  E.~E.\ 1979, \apj, 231, 438

\bibitem[Elias-Rosa et al.(2008)]{eliasrosa08} Elias-Rosa, N., et
  al.\ 2008, \mnras, 384, 107

\bibitem[Ferland et al.(1998)]{ferland98} Ferland, G.~J., Korista,
  K.~T., Verner, D.~A., Ferguson, J.~W., Kingdon, J.~B., \& Verner,
  E.~M.\ 1998, \pasp, 110, 761

\bibitem[Fesen et al.(1989){Fesen, Saken, \& Hamilton}]{fesen89}
  Fesen, R.~A., Saken, J.~M., \& Hamilton, A.~J.~S.\ 1989, \apjl, 341,
  L55

\bibitem[Filippenko(1997)]{filippenko97} Filippenko, A.~V.\ 1997,
  \araa, 35, 309

\bibitem[Filippenko(2005)]{filippenko05} ------ 2005, in White Dwarfs:
  Cosmological and Galactic Probes, ed. E. M. Sion, S. Vennes, \&
  H. L. Shipman (Dordrecht: Springer), 97

\bibitem[Filippenko et al.(2001)]{filippenko01} Filippenko, A.~V., Li,
  W.~D., Treffers, R.~R., \& Modjaz, M.\ 2001, in ASP Conf. Ser. 246, Small
  Telescope Astronomy on Global Scales, ed. W. P. Chen, C. Lemme, 
  \& B. Paczy\'{n}ski  (San Francisco: ASP), 121

\bibitem[Filippenko et al.(2007)]{cbet1101} Filippenko, A.~V.,
  Silverman, J.~M., Foley, R.~J., Modjaz, M., Papovich, C., Willmer,
  C.~N.~A., Blondin, S., \& Brown, W.\ 2007, CBET 1101, 1

\bibitem[Foley et al.(2003)]{foley03} Foley, R.~J., et al.\ 2003,
  \pasp, 115, 1220

\bibitem[Gonz{\' a}lez-Hern{\'a}ndez et al.(2009)]{gh08}
  Hern{\'a}ndez, J.~I.~G., Ruiz-Lapuente, P., Filippenko, A.~V.,
  Foley, R.~J., Gal-Yam, A., \& Simon, J.~D.\ 2009, \apj, 691, 1

\bibitem[Hamuy et al.(1996)]{hamuy96} Hamuy, M., Phillips, M.~M.,
  Suntzeff, N.~B., Schommer, R.~A., Maza, J., \& Aviles, R.\ 1996,
  \aj, 112, 2391

\bibitem[Hamuy et al.(2003)]{hamuy03} Hamuy, M., et al.\ 2003, \nat,
  424, 651

\bibitem[Herbig(1968)]{herbig68} Herbig, G.~H.\ 1968, \zap, 68, 243

\bibitem[Hicken et al.(2009)]{hicken09} Hicken, M., Wood-Vasey, W.~M.,
  Blondin, S., Challis, P., Jha, S., Kelly, P.~L., Rest, A., \&
  Kirshner, R.~P.\ 2009, arXiv:0901.4804

\bibitem[Horne(1986)]{horne86} Horne, K.\ 1986, \pasp, 98, 609 

\bibitem[Howard et al.(1963){Howard, Wentzel, \& McGee}]{howard63}
  Howard, W.~E., III, Wentzel, D.~G., \& McGee, R.~X.\ 1963, \apj,
  138, 988

\bibitem[Howell(2001)]{howell01} Howell, D.~A.\ 2001, \apjl, 554, L193

\bibitem[Howell et al.(2005)]{howell05} Howell, D.~A., et al.\ 2005,
  \apj, 634, 1190

\bibitem[Hughes et al.(2007)]{hughes07} Hughes, J.~P., Chugai, N.,
  Chevalier, R., Lundqvist, P., \& Schlegel, E.\ 2007, \apj, 670, 1260

\bibitem[Iben \& Tutukov(1984)]{it84} Iben, I., Jr., \& Tutukov,
  A.~V.\ 1984, \apjs, 54, 335

\bibitem[Ihara et al.(2007)]{ihara07} Ihara, Y., Ozaki, J., Doi, M.,
  Shigeyama, T., Kashikawa, N., Komiyama, K., \& Hattori, T.\ 2007,
  \pasj, 59, 811

\bibitem[Immler et al.(2006)]{immler06} Immler, S., et al.\ 2006,
  \apjl, 648, L119

\bibitem[Jha et al.(2007){Jha, Riess, \& Kirshner}]{jrk07} Jha, S.,
  Riess, A.~G., \& Kirshner, R.~P.\ 2007, \apj, 659, 122

\bibitem[Kennicutt et al.(2003){Kennicutt, Bresolin, \&
    Garnett}]{kennicutt03} Kennicutt, R.~C., Jr., Bresolin, F., \&
  Garnett, D.~R.\ 2003, \apj, 591, 801

\bibitem[Kerzendorf et al.(2009)]{kerz09} Kerzendorf, W.~E., Schmidt,
  B.~P., Asplund, M., Nomoto, K., Podsiadlowski, P., Frebel, A.,
  Fesen, R.~A., \& Yong, D.\ 2009, \apj, in press (preprint at ArXiv
  e-prints, 906, arXiv:0906.0982)

\bibitem[Koribalski et al.(2004)]{koribalski04} Koribalski, B.~S., et
  al.\ 2004, \aj, 128, 16

\bibitem[Krause et al.(2008)]{krause08} Krause, O., Tanaka, M., Usuda,
  T., Hattori, T., Goto, M., Birkmann, S., \& Nomoto, K.\ 2008, \nat,
  456, 617

\bibitem[Leibundgut(2004)]{leibundgut04} Leibundgut, B.\ 2004, \apss,
  290, 29

\bibitem[Leonard(2007)]{leonard07} Leonard, D.~C.\ 2007, \apj, 
670, 1275 

\bibitem[Li et al.(2000)]{li00} Li, W.~D., et al.\ 2000, in AIP
  Conf. Ser. 522, Cosmic Explosions, ed. S. S. Holt \& W. W. Zhang
  (New York: AIP), 103

\bibitem[Li et al.(2001)]{li01a} Li, W., Filippenko, A.~V., Treffers,
  R.~R., Riess, A.~G., Hu, J., \& Qiu, Y.\ 2001, \apj, 546, 734

\bibitem[Li et al.(2001)]{li01b} Li, W., et al.\ 2001, \pasp, 113,
  1178

\bibitem[Li et al.(2003)]{li03} ------\ 2003, \pasp, 115, 453

\bibitem[Magrini et al.(2007)]{magrini07} Magrini, L., V{\'{\i}}lchez,
  J.~M., Mampaso, A., Corradi, R.~L.~M., \& Leisy, P.\ 2007, \aap,
  470, 865

\bibitem[Mannucci et al.(2005)]{mannucci05} Mannucci, F., Della Valle,
  M., Panagia, N., Cappellaro, E., Cresci, G., Maiolino, R.,
  Petrosian, A., \& Turatto, M.\ 2005, \aap, 433, 807

\bibitem[Mannucci et al.(2006){Mannucci, Della Valle, \&
    Panagia}]{mannucci06} Mannucci, F., Della Valle, M., \& Panagia,
  N.\ 2006, \mnras, 370, 773

\bibitem[Marietta et al.(2000){Marietta, Burrows, \&
    Fryxell}]{marietta00} Marietta, E., Burrows, A., \& Fryxell,
  B.\ 2000, \apjs, 128, 615

\bibitem[Matheson et al.(2000)]{matheson00} Matheson, T., Filippenko,
  A.~V., Ho, L.~C., Barth, A.~J., \& Leonard, D.~C.\ 2000, \aj, 120,
  1499

\bibitem[Matheson et al.(2001)]{matheson01} Matheson, T., Filippenko,
  A.~V., Li, W., Leonard, D.~C., \& Shields, J.~C.\ 2001, \aj, 121,
  1648

\bibitem[Mattila et al.(2005)]{mattila05} Mattila, S., Lundqvist, P.,
  Sollerman, J., Kozma, C., Baron, E., Fransson, C., Leibundgut, B.,
  \& Nomoto, K.\ 2005, \aap, 443, 649

\bibitem[Mazzali et al.(1997)]{mazzali97} Mazzali, P.~A., Chugai, N.,
  Turatto, M., Lucy, L.~B., Danziger, I.~J., Cappellaro, E., della
  Valle, M., \& Benetti, S.\ 1997, \mnras, 284, 151

\bibitem[Mazzali et al.(2005a)]{mazzali05a} Mazzali, P.~A., et
  al.\ 2005a, \apjl, 623, L37

\bibitem[Mazzali et al.(2005b)]{mazzali05b} Mazzali, P.~A., Benetti, S.,
  Stehle, M., Branch, D., Deng, J., Maeda, K., Nomoto, K., \& Hamuy,
  M.\ 2005b, \mnras, 357, 200

\bibitem[Miller \& Stone(1993)]{miller93} Miller, J. S., \& Stone,
   R. P. S. 1993, Lick Obs. Tech. Rep. 66

\bibitem[Monard et al.(2007){Monard, Yamaoka, \& Itagaki}]{cbet1100}
  Monard, L.~A.~G., Yamaoka, H., \& Itagaki, K.\ 2007, CBET 1100, 1

\bibitem[Moore et al.(1966){Moore, Minnaert, \& Houtgast}]{moore66}
  Moore, C.~E., Minnaert, M.~G.~J., \& Houtgast, J.\ 1966, National
  Bureau of Standards Monograph, Washington: US Government Printing
  Office (USGPO)

\bibitem[Morton(1975)]{morton75} Morton, D.~C.\ 1975, \apj, 197, 85

\bibitem[Murray et al.(2007)]{murray07} Murray, N., Martin, C.~L.,
  Quataert, E., \& Thompson, T.~A.\ 2007, \apj, 660, 211

\bibitem[Nugent et al.(1995)]{nugent95} Nugent, P., Baron, E.,
  Hauschildt, P.~H., \& Branch, D.\ 1995, \apjl, 441, L33

\bibitem[O'Brien et al.(2006)]{obrien06} O'Brien, T.~J., et al.\ 2006,
  \nat, 442, 279

\bibitem[Oemler \& Tinsley(1979)]{ot79} Oemler, A., Jr., \& Tinsley,
  B.~M.\ 1979, \aj, 84, 985

\bibitem[Oke et al.(1995)]{oke95} Oke, J.~B., et al.\ 1995, \pasp,
  107, 375

\bibitem[Pakmor et al.(2008)]{pakmor08} Pakmor, R., R{\"o}pke, F.~K.,
  Weiss, A., \& Hillebrandt, W.\ 2008, \aap, 489, 943

\bibitem[Panagia et al.(2006)]{panagia06} Panagia, N., Van Dyk, S.~D.,
  Weiler, K.~W., Sramek, R.~A., Stockdale, C.~J., \& Murata,
  K.~P.\ 2006, \apj, 646, 369

\bibitem[Patat et al.(2007a)]{patat} Patat, F., et al. 2007a, Science,
  317, 924

\bibitem[Patat et al.(2007b)]{patat07b} ------\ 2007b, \aap, 474, 931

\bibitem[Poole et al.(2008)]{swiftcal} Poole, T.~S., et al.\ 2008,
  \mnras, 383, 627

\bibitem[Prieto et al.(2007)]{prieto07} Prieto, J.~L., et al.\ 2007,
  submitted to \aj\ (preprint at ArXiv e-prints, 706, arXiv:0706.4088)

\bibitem[Quimby et al.(2007){Quimby, H{\"o}flich, \&
  Wheeler}]{quimby07} Quimby, R., H{\"o}flich, P., \& Wheeler,
  J.~C.\ 2007, \apj, 666, 1083

\bibitem[Riess et al.(2007)]{riess07} Riess, A.~G., et al.\ 2007,
  \apj, 659, 98

\bibitem[Ruiz-Lapuente(2004)]{rl04a} Ruiz-Lapuente, P.\ 2004, \apj,
  612, 357

\bibitem[Ruiz-Lapuente et al.(2004)]{rl04} Ruiz-Lapuente, P., et
  al.\ 2004, \nat, 431, 1069

\bibitem[Sauer et al.(2008)]{sauer08} Sauer, D.~N., et al.\ 2008,
  \mnras, 391, 1605

\bibitem[Scannapieco \& Bildsten(2005)]{sb05} Scannapieco, E., \&
  Bildsten, L.\ 2005, \apjl, 629, L85

\bibitem[Schlegel et al.(1998){Schlegel, Finkbeiner, \& Davis}]{sfd98}
  Schlegel, D.~J., Finkbeiner, D.~P., \& Davis, M.\ 1998, \apj, 500,
  525

\bibitem[Simon et al.(2007)]{simon07} Simon, J.~D., et al.\ 2007,
  \apjl, 671, L25

\bibitem[Spitzer(1954)]{spitzer54} Spitzer, L.~J.\ 1954, \apj, 120, 1

\bibitem[Spitzer(1978)]{spitzer78} ------\ 1978, Physical
  Processes in the Interstellar Medium (New York: New York
  Wiley-Interscience)

\bibitem[Stetson(1987)]{stetson87} Stetson, P.~B.\ 1987, \pasp, 99, 191

\bibitem[Sullivan et al.(2006)]{sullivan06} Sullivan, M., et
  al.\ 2006, \apj, 648, 868

\bibitem[Tanaka et al.(2008)]{tanaka08} Tanaka, M., et al.\ 2008,
  \apj, 677, 448

\bibitem[Totani et al.(2008)]{totani08} Totani, T., Morokuma, T., Oda,
  T., Doi, M., \& Yasuda, N.\ 2008, \pasj, 60, 1327

\bibitem[Trundle et al.(2008)]{trundle08} Trundle, C.,
  Kotak, R., Vink, J.~S., \& Meikle, W.~P.~S.\ 2008, \aap, 483, L47

\bibitem[Tull(1998)]{hrs} Tull, R.~G.\ 1998, \procspie, 3355, 387

\bibitem[van Hoof et al.(2004)]{vanhoof04} van Hoof, P.~A.~M.,
  Weingartner, J.~C., Martin, P.~G., Volk, K., \& Ferland,
  G.~J.\ 2004, \mnras, 350, 1330

\bibitem[Verner et al.(1996)]{verner96} Verner, D.~A., Ferland, G.~J.,
  Korista, K.~T., \& Yakovlev, D.~G.\ 1996, \apj, 465, 487

\bibitem[Vogt et al.(1994)]{vogt94} Vogt, S.~S., et al.\ 1994,
  \procspie, 2198, 362

\bibitem[Wade \& Horne(1988)]{wh88} Wade, R.~A., \& Horne, K.\ 1988,
  \apj, 324, 411

\bibitem[Wang et al.(2007)]{wang07} Wang, L., Baade, D., \& Patat,
  F.\ 2007, Science, 315, 212

\bibitem[Wang et al.(2006)]{wang06} Wang, L., Baade, D., Patat, F., \&
  Wheeler, J.~C.\ 2006, CBET 396, 2

\bibitem[Wang \& Wheeler(2008)]{ww08} Wang, L., \& Wheeler,
  J.~C.\ 2008, \araa, 46, 433

\bibitem[Wang et al.(2003)]{wang03} Wang, S., et al.\ 2003,
  \procspie, 4841, 1145

\bibitem[Wang et al.(2008b)]{wang08b} Wang, X., Li, W.,
  Filippenko, A.~V., Foley, R.~J., Smith, N., \& Wang, L.\ 2008b, \apj,
  677, 1060

\bibitem[Wang et al.(2008a)]{wang08a} Wang, X., et al.\ 2008a, \apj, 675,
  626

\bibitem[Wang et al.(2009)]{wang09} ------\ 2009, \apj, 697, 380 

\bibitem[Webbink(1984)]{webbink84} Webbink, R.~F.\ 1984, \apj, 277,
  355

\bibitem[Welty \& Fitzpatrick(2001)]{wf01} Welty, D.~E., \&
  Fitzpatrick, E.~L.\ 2001, \apj, 551, L175

\bibitem[Whelan \& Iben(1973)]{wi73} Whelan, J., \& Iben, I.~J.\ 1973,
  \apj, 186, 1007

\bibitem[Williams et al.(2008)]{williams08} Williams, R., Mason, E.,
  Della Valle, M., \& Ederoclite, A.\ 2008, \apj, 685, 451

\bibitem[Wood-Vasey et al.(2007)]{wv07} Wood-Vasey, W.~M., et
  al.\ 2007, \apj, 666, 694

\end{thebibliography}
\end{document}